

\input uiucmac.tex
\font\bfgreek=cmmib10 scaled \magstep1
\def\bmath{\fam1\bfgreek\textfont1=\bfgreek}

\def\Ai{\hbox{Ai}}
\def\Bi{\hbox{Bi}}

\PHYSREV

\sequentialequations
\tolerance 2000
\nopubblock
\titlepage
%
%
\singlespace
\title{\bf The Renormalization Group and Singular Perturbations:
Multiple-Scales, Boundary Layers and Reductive Perturbation Theory}
\author{Lin-Yuan Chen\rlap,$^{1,2}$ Nigel Goldenfeld$^{1}$ and Y. Oono$^{1}$}
\address{$^1$Department of Physics, Materials Research
Laboratory, and Beckman Institute, 1110 West Green Street, University of
Illinois at Urbana-Champaign, Urbana, IL 61801-3080, U. S. A.}
\address{$^2$Institute for Theoretical Physics, University of
California, Santa Barbara, CA 93106-4030, U. S. A.}
\smallskip
%
\abstract
\noindent

Perturbative renormalization group theory is developed as a unified tool
for global asymptotic analysis.  With numerous examples, we illustrate
its application to ordinary differential equation problems involving
multiple scales, boundary layers with technically difficult asymptotic
matching, and WKB analysis.  In contrast to conventional methods, the
renormalization group approach requires neither {\it ad hoc\/}
assumptions about the structure of perturbation series nor the use of
asymptotic matching. Our renormalization group approach provides
approximate solutions which are practically superior to those obtained
conventionally, although the latter can be reproduced, if desired, by
appropriate expansion of the renormalization group approximant.  We show
that the renormalization group equation may be interpreted as an
amplitude equation, and from this point of view develop reductive
perturbation theory for partial differential equations describing
spatially-extended systems near bifurcation points, deriving both
amplitude equations and the center manifold.

\medskip
\noindent
Pacs numbers: 47.20.Ky, 02.30.Mv, 64.60.Ak
\endpage

{\bf\chapter{Introduction}}

Asymptotic and perturbative analysis has played a significant role in
applied mathematics and theoretical physics. In many cases, regular
perturbation methods are not applicable, and various singular
perturbation techniques must be used\rlap.\REFS\bender{C.\ M.\ Bender
and S.\ A.\ Orszag, {\sl Advanced Mathematical Methods for Scientists
and Engineers} (McGraw-Hill,1978)}\REFSCON\pertbooks{J.\ Kevorkian and
J.\ D.\ Cole, {\sl Perturbation Methods in Applied Mathematics}
(Springer, New York,1981); A. Nayfeh, {\sl Perturbation Methods} (John
Wiley \& Sons, New York, 1973); D.\ R.\ Smith, {\sl Singular
Perturbation Theory} (Cambridge UP., Cambridge, 1985); 
S.\ A.\ Lomov, {\sl Introduction to the General Theory of Singular
Perturbation} (Am.\ Math.\ Soc., Providence,
1992).}\REFSCON\murdock{J.\ A.\ Murdock,
{\sl Perturbations Theory and Methods} (Wiley, New York, 1991).}
\REFSCON\robert{A.\ J.\ Roberts, \journal SIAM J.  Math. Anal.
&16&1243(85).}\REFSCON\hinch{E.\ J.\ Hinch, {\sl Perturbation Methods}
(Cambridge UP., Cambridge, 1991).}  \REFSCON\carr{J.\ Carr, {\sl
Applications of Center Manifold Theory} (Springer, Berlin, 1981);
J.\ Carr and R.\ M.\ Muncaster, \journal J.  Diff. Eq.
&50&260(83).}\refsend\ Examples of widely-used techniques for ordinary
differential equations (ODEs) include\refmark{\bender,\pertbooks} the
methods of multiple scales, boundary layers or asymptotic
matching, WKB, stretched coordinates, averaging, the method of
reconstitution\rlap,\refmark{\robert} and center manifold 
theory\rlap.\refmark{\carr}  Although these methods are well known, each
has its own drawbacks, preventing mechanical (or algorithmic)
application.  Indeed, it is probably fair to say that the practice of
asymptotic analysis is something of an art.

Multiple-scales analysis has proven to be a particularly useful tool for
constructing uniform or global approximate solutions for both small and
large values of independent variables.  In this method a set of scaled
variables, which are regarded as independent variables (although they
are ultimately related to one another), is introduced to remove all
secular terms.  The choice of the set is, in some cases, nontrivial, and
may only be justified {\it post hoc}.  Nevertheless, this method is
usually considered the most general, subsuming the others mentioned
below.

Differential equations whose highest order derivatives are multiplied by
a small parameter $\epsilon$ often yield solutions with narrow regions
of rapid variation, known as boundary layers.  Boundary layer techniques
can be applied if the thickness of these regions tends to zero as
$\epsilon\rightarrow 0$; otherwise, WKB must be used.  The limitation of
WKB is that it applies to linear equations only.  Although boundary
layer methods apply to nonlinear as well as to linear problems, the
determination of the expansion parameter can be subtle.  Furthermore,
matching of outer and inner expansions via intermediate expansions is
required, sometimes involving delicate arguments that are difficult to
perform mechanically.

Another class of related problems concerns partial differential
equations (PDEs) describing nonequilibrium, spatially-extended systems
near bifurcation points.   Such systems often exhibit spatial-temporal
patterns modulated by an envelope function (or amplitude) which varies
slowly compared with the pattern itself.  Extracting the long
wavelength, slow timescale behavior of such systems is the task of
reductive perturbation methods\rlap,\Ref\reductive{T.\ Taniuti and
C.\ C.\ Wei, \journal J. Phys.  Soc. Jpn. &24&941(68); A.\ C.\ Newell
and J.\ A.\ Whitehead, \journal J. Fluid Mech. &38&279(69);
Y.\ Kuramoto, {\sl Chemical Oscillations, Waves, and Turbulence}
(Springer, Berlin, 1984).} which are themselves related to
multiple-scales analysis.

The purpose of this paper is to present a unified, and
physically-motivated approach to these classes of problems, based upon
the renormalization group (RG).  The essence of the renormalization
group method is to extract structurally stable features of a system
which are insensitive to details\rlap.\REFS\oonoi{Y.\ Oono,
Adv.\ Chem.\ Phys.\ {\bf 61}, Chapter 5 (edited by I.\ Prigogine and
S.\ A.\ Rice, Wiley, 1985).}\REFSCON\oonoii{ Y.\ Oono,  \journal Butsuri
&42&311(1987).}\REFSCON\nigeli{N.\ D.\ Goldenfeld, {\sl Lectures on
Phase Transitions and the Renormalization Group} (Addison-Wesley,
Reading, Mass., 1992).}\REFSCON\kawasakifest{L.\ Y.\ Chen,
N.\ Goldenfeld, Y.\ Oono and G.\ Paquette, \journal Physica A
&204&111(1994); G.\ Paquette, L.\ Y.\ Chen, N.\ Goldenfeld, and
Y.\ Oono, \journal Phys.  Rev. Lett. &72&76(1994); G. C. Paquette and
Y.  Oono, \journal Phys. Rev. E &49&2368(94); L.-Y. Chen, N. D.
Goldenfeld, and Y. Oono, \journal Phys. Rev. E
&49&4502(94).}\refsend\ For example, field theories, critical phenomena,
polymers and other statistical mechanical systems exhibit universal
scaling functions and critical exponents in the limit
$\Lambda/\xi\rightarrow 0$, where $\Lambda$ is some ultra-violet cut-off
and $\xi$ is the (temperature-dependent) correlation length. The
renormalization group is the principal tool with which to elucidate this
universal behavior and is properly regarded as a means of asymptotic
analysis.
 
The usefulness of this point of view has been amply
demonstrated\REFS\rgpde{N.\ Goldenfeld, O.\ Martin and Y.\ Oono,
\journal J. Sci.  Comp.&4&355(89); N.\ Goldenfeld, O.\ Martin, Y.\ Oono
and F.\ Liu, \journal Phys. Rev.  Lett.
&64&1361(90).}\REFSCON\bricmont{J.\ Bricmont and A.\ Kupiainen, \journal
Commun.  Math. Phys.  &150&193(92); J.\ Bricmont, A.\ Kupiainen and
G.\ Lin, \journal Commun. Pure Appl. Math. &47&893(1994); J. Bricmont
and A. Kupiainen, {\it Renormalizing Partial Differential Equations},
preprint {\tt chao-dyn/9411015}; J. Bricmont,
A. Kupiainen and J. Xin, {\it Global Large Time Self-similarity of
a Thermal-Diffusive Combustion System with Critical Nonlinearity},
preprint {\tt chao-dyn/9501012}.}\refsend\ by
the relationship between the renormalization group and intermediate
asymptotics\rlap.\Ref\barenblatt{G.\ I.\ Barenblatt, {\sl Similarity,
Self-Similarity, and Intermediate Asymptotics} (Consultant Bureau, New
York, 1979).}  In particular, the large-time asymptotic behavior of
certain initial-value problems is given by a similarity solution of the
governing PDE, where the similarity variable contains anomalous
exponents which may not be determined {\it a priori\/} by elementary
dimensional considerations.  Nevertheless, renormalized perturbation
theory combined with the renormalization group, gives an expansion for
the anomalous exponents and the solution\rlap.\Ref\pedagog{For a
pedagogical account of the relationship between the anomalous dimensions
of quantum field theory and those of partial differential equations, see
ref. \nigeli, chapter 10.}

The similarities between renormalization group and singular perturbation
methods extend also to technical details: both perturbative
renormalization group and conventional singular perturbation methods
remove secular or divergent terms from perturbation series. These formal
similarities invite a natural question: what is the relation, if any,
between conventional asymptotic methods and the renormalization group?

In this paper, which is an extended version of our preliminary
report\rlap,\Ref\sprl{L.-Y.\ Chen, N.\ Goldenfeld, and Y. Oono, \journal
Phys. Rev. Lett. &73&1311(94).} we demonstrate that singular
perturbation methods may be naturally understood as renormalized
perturbation theory, and that amplitude equations obtainable by
reductive perturbation methods may be derived as renormalization group
equations.

Our studies indicate that the renormalization group method may have
several advantages compared with conventional methods.  Although we
recognize that our analysis is at the formal, heuristic level, we
suggest that a more careful mathematical analysis would be worthwhile,
given the potential usefulness of our central claim.

One advantage of the renormalization group method is that the starting
point is a straightforward naive perturbation expansion, for which very
little {\it a priori\/} knowledge is required.  That is, one does not
need to guess or otherwise introduce unexpected fractional power laws or
logarithmic functions of $\epsilon$ in an {\it ad hoc\/} manner.  It
seems that these $\epsilon$-dependent space/time scales arise naturally
during the analysis.

We will show that the renormalization group approach sometimes seems to
be more efficient and accurate in practice than standard methods in
extracting global information from the perturbation expansion.  Standard
methods often attempt to represent an asymptotic solution in terms of
asymptotic sequences of a few simple functions of the expansion
parameter, such as exp, log, powers and so on.  The renormalization
group can generate its own problem-adapted asymptotic sequence without
matching: in the examples given in section 4, these turn out to be
complicated functions conveniently defined by an integral
representation.  For small $\epsilon$, this asymptotic sequence can be
expanded to reproduce the solutions conventionally obtained by
asymptotic matching, although in the examples that we have studied so
far, the conventional approximant is practically inferior to the one
obtained by the RG.  In switchback problems, the RG perturbation series
may need to be carried out to higher than lowest order, then expanded in
$\epsilon$, in order to reproduce the (inferior) conventional result.

A related advantage of the renormalization group seems to be the lack of
necessity to perform asymptotic matching.  To illustrate this assertion,
in section 3 we solve several ODEs with boundary layers, and in section
4 we address the difficult technical problem of switchback terms.

The renormalization group methods for partial differential equations
such as the Barenblatt
equation\rlap,\refmark{\rgpde,\bricmont,\barenblatt} and front
propagation problems in reaction-diffusion
equations\rlap,\refmark{\kawasakifest} are, in retrospect, examples of
the general approach discussed in this paper.  We emphasize that our
renormalization group method has no connection with the so-called method
of renormalization or uniformization\refmark{\bender} in the
conventional perturbation literature; the latter is a mere variant of
the method of stretched coordinates, and of narrow limited use.

Lastly, we wish to point out that recently, a method utilizing an
invariance condition in the solution of multiple-scale singular
perturbation problems was proposed independently by
Woodruff\rlap,\Ref\woodruff{S. L. Woodruff, \journal Studies in Applied
Mathematics &90&225(93).}  based on ideas related to the renormalization
group.  In addition, Kunihiro\Ref\kunihiro{T. Kunihiro, {\sl A
geometrical formulation of the renormalization group method for global
analysis}, preprint available from archive {\tt hep-th/9505166}.} has
shown that the renormalization group method that we proposed in ref.
\sprl\ may be interpreted geometrically in terms of the classical theory
of envelopes.  

The outline of this paper is as follows.  In Section 2, we discuss the
general relation between  multiple-scale analysis and renormalization
group. In Section 3, we show how boundary layer and WKB problems can be
solved using the renormalization group.  In Section 4, we demonstrate
with several examples that the renormalization group approach has
technical advantages to conventional asymptotic methods.  In Section 5,
the renormalization group is applied as a reductive perturbation tool to
the derivation of global slow motion equations for partial differential
equations.  Center manifold theory is also briefly considered from the
same point of view.  We conclude in Section 6.

\medskip

{\bf\chapter{Multiple Scale Theory and RG}}
\smallskip

In this section, we show that multiple-scale analysis is equivalent to
the RG, and that the solvability condition used in multiple scales to
remove the secular divergences is equivalent to the physical assumption
of renormalizability in RG theory.

\medskip

{\bf \section{Rayleigh Equation}}
\smallskip

The example we consider below is the Rayleigh
equation\rlap,\Ref\benderrayleigh{See ref. \bender, page 554.} closely
related to the Van der Pol oscillator:
$$
\frac{d^2 y}{d t^2}+y=\epsilon\Big\{\frac{d y}{d t}-\frac{1}{3}
\left(\frac{d y}{d t}\right)^3\Big\}.\eqn\twelve
$$
It is known that the method of uniformization or
renormalization\refmark{\bender} fails here, and this example is a
textbook illustration of multiple scales analysis. We show here that
from only the simple-minded straightforward expansion, not only is the
RG capable of identifying automatically all different multiple scales
required by multiple scales analysis, but also produces a uniformly
valid asymptotic solution without encountering the ambiguity which often
plagues higher order calculations in multiple scales analysis.

A naive expansion $y=y_0+\epsilon y_1+\epsilon^2 y_2+\cdots$ gives
$$
\eqalign{
y(t)=&R_0 \sin(t+\Theta_0)+\epsilon \Big\{-\frac{R_0^3}{96}\cos(t+\Theta_0)\cr
 &+ \frac{R_0}{2}\left(1-\frac{R_0^2}{4}\right)(t-t_0)\sin(t+\Theta_0)
+\frac{R_0^3}{96}\cos3(t+\Theta_0)\Big\}+O(\epsilon^2),\cr}\eqn\thirteen
$$
where $R_0,\Theta_0$ are constants determined by the initial conditions
at arbitrary $t=t_0$.  This naive perturbation theory breaks down when
$\epsilon (t-t_0) > 1$ because of the secular terms.  The arbitrary time
$t_0$ may be interpreted as the (logarithm of the) ultraviolet cutoff in
the usual field theory.\refmark{\kawasakifest} To regularize the
perturbation series, we introduce an arbitrary time $\tau$, split
$t-t_0$ as $t-\tau + \tau - t_0$, and absorb the terms containing
$\tau-t_0$ into the renormalized counterparts $R$ and $\Theta$ of $R_0$
and $\Theta_0$, respectively.  This is allowed because $R_0$ and
$\Theta_0$ are no longer constants of motion in the presence of the
nonlinear perturbation.

We introduce a multiplicative renormalization constant
$Z_1=1+\sum_{1}^{\infty} a_n \epsilon^n$ and an additive one
$Z_2=\sum_{1}^{\infty} b_n \epsilon^n$ such that $R_0(t_0)=Z_1(t_0,\tau)
R(\tau)$ and $\Theta_0(t_0)=\Theta(\tau) +Z_2(t_0,\tau)$.  The
coefficients $a_n$ and $b_n$ ($n\ge 1$) are chosen order by order in
$\epsilon$ to eliminate the terms containing $\tau-t_0$ as in the
standard RG\rlap.\REFS\wilson{K.G. Wilson, \journal Phys. Rev. B
&4&3174(71); \journal ibid. &4&3184(71).}\REFSCON\stueck{ E.
Stueckelberg, and A. Petermann, \journal Helv. Phys. Acta &26&499(53).}
\REFSCON\gellmann{M. Gell-Mann and F.E.  Low, \journal Phys. Rev.
&95&1300(54); see also the presentations by K. Wilson, \journal Phys.
Rev. &179&1499(69)\ and  by J.D. Bjorken and S.D. Drell, {\sl
Relativistic Quantum Fields} (McGraw-Hill, New York,
1966).}\REFSCON\bs{N.N. Bogoliubov and D.V. Shirkov, \journal Usp. Fiz.
Nauk &55&149(55); \journal ibid. &57&3(55); \journal JETP &30&77(56)
[\journal Sov. Phys. JETP &3&57(56)]; for a pedagogical discussion of
the connection with Lie groups, see N.N. Bogoliubov and D.V. Shirkov,
{\sl Introduction to the Theory of Quantized Fields} (3rd edition)
(Wiley, New York, 1980).}\REFSCON\zinn{J.  Zinn-Justin, {\sl Quantum
Field Theory and Critical Phenomena} (Clarendon, Oxford,
1989).}\REFSCON\amit{D. J.  Amit, {\sl Field Theory, the Renormalization
Group and Critical Phenomena} (World Scientific, Singapore,
1984).}\refsend The choice $a_1=-(1/2)(1-R^2/4)(\tau-t_0)$, $b_1=0$
removes the secular terms to order $\epsilon$, and we obtain the
following renormalized perturbation result\Ref\rgconst{$Z_1$ may depend
upon $R$, because $R$ is dimensionless.  This is analogous to the
renormalization of a dimensionless coupling constant in field theory.}
$$\eqalign{
y(t)=&\Big\{R+\epsilon \frac{R}{2}\left(1-\frac{R^2}{4}\right)
(t-\tau)\Big\}\sin(t+\Theta)\cr
&-\epsilon \frac{1}{96} R^3 \cos(t+\Theta)
+\epsilon \frac{R^3}{96} \cos3(t+\Theta)+O(\epsilon^2),\cr}\eqn\fourteen$$
where $R,\Theta$ are now functions of $\tau$.  Since $\tau$ does not
appear in the original problem, the solution should not depend on
$\tau$.  Therefore, $(\partial y/\partial \tau)_t= 0$ for any $t$.
This is the RG equation, which in this case consists of two independent
equations
$$\frac{d R}{d \tau}=\epsilon \frac{1}{2}R\left(1-
\frac{1}{4}R^2\right)+O(\epsilon^2)
,\quad \frac{d \Theta}{d \tau}=O(\epsilon^2).\eqn\fifteen$$
Solving \fifteen\  and equating $\tau$ and $t$ eliminates the secular
term, we get
$$R(t)=R(0)/\sqrt{e^{-\epsilon t}+\frac{1}{4}R(0)^2(1-e^{-\epsilon t})}
+O(\epsilon^2 t), \quad \Theta(t)=\Theta(0)+O(\epsilon^2 t),
\eqn\sixteen$$
where $R(0),\Theta(0)$ are constants to be determined by the initial
condition. Assuming the initial condition $y(0)=0,y'(0)=2a$, we find
$R(0)=2a,\Theta(0)=0$, and the final uniformly valid result reads
$$y(t)=R(t)\sin(t)+\frac{\epsilon}{96} R(t)^3
\left\{\cos(3t)-\cos(t)\right\}+O(\epsilon^2),
\eqn\seventeen$$
which approaches a limit circle of radius 2 as $t\rightarrow \infty$.

The second order RG calculation shows the assumption of perturbative
renormalizability is consistent and no ambiguity arises at all. The
corresponding amplitude and phase equation to order $O(\epsilon^3)$ are
$$
\eqalign{
\frac{d R}{d t}=&\epsilon \frac{1}{2}R\left(1-\frac{1}{4}R^2\right)
+O(\epsilon^3), \cr
\frac{d \Theta}{d t}=&-\frac{\epsilon^2}{8}
\left(1-\frac{R^4}{32}\right)+O(\epsilon^3),\cr}
\eqn\eightteen$$
from which the multiple time scales $T_1=\epsilon t$,
$T_2=\epsilon^2 t, \cdots$ used in multiple scales analysis appear naturally (although the
RG does not require such identifications).  When $R=2$,
\eightteen\ reduces to
$$\frac{d R}{d t}=0+O(\epsilon^3),\quad\quad \frac{d \Theta}{d
t}=-\frac{1}{16} \epsilon^2+O(\epsilon^3).\eqn\eightteencom$$

In this simple example, it was straightforward to determine the multiple
time scales.  However, it is well known that in many cases, within
multiple scales analysis hidden intermediate scales must be included in
the perturbation expansion so as to obtain the correct result.  In the
next example, will show that the RG method is a more straightforward but
secure way to determine multiple slow time scales than the multiple
scales method.

{\bf\section{Mathieu Equation}}
\smallskip

The second illustrative example we examine using RG is the Mathieu
equation\Ref\bendermathieu{See ref. \bender, page 560.}
$$\frac{d^2 y}{d
t^2}+(a+2\epsilon \cos t)y=0,\eqn\mathieu$$ where $a$ and $\epsilon$
are parameters.

The Floquet theory of linear periodic differential
equations\refmark{\bender} predicts that in the $(a,\epsilon)$ plane
there are some regions where the solutions to \mathieu\ remain bounded
for all $t$ and stable, and others where the solutions are unstable.
Perturbative investigation shows that for sufficiently small $\epsilon$,
all solutions $y(t)$ are stable for $a>0, a\not=n^2/4$, $n=0, 1, 2,
\cdots$.  Without loss of generality, we investigate the stability of
solutions near $a=1/4$ and $\epsilon=0$ to find the stability boundary
in the $(a,\epsilon)$ plane. We treat the boundary curve $a$ as a
function of $\epsilon$ and expand $a$ in powers of $\epsilon$:
$a(\epsilon) =1/4 + a_1 \epsilon +a_2 \epsilon^2 +\cdots$. It is our
goal to determine values of $a_1,a_2,\cdots$ perturbatively.
Multiple-scale analysis can be applied to this problem, and the
coefficients $a_1=1, a_2=-1/2$ are determined. However, it turns out
that the introduction of multiple time scales $\tau_1 =\epsilon t$, $
\tau_2 =\epsilon^2 t,\cdots$ is not sufficient to determine the second
order coefficient $a_2$ even after the first order coefficient $a_1$ is
set to $1$. Through careful analysis, it is found that a new hidden time
scale $\sigma=\epsilon^{3/2} t$ must be introduced into the problem, and
the perturbative expansion must be done in powers of $\epsilon^{1/2}$,
rather than the original expansion in powers of $\epsilon$. It is
necessary to go to the fourth order in powers of $\epsilon^{1/2}$ to
determine $a_2$. Thus, the procedure required to determine all necessary
time scales is not mechanical: if any hidden scales are omitted or
cannot be  determined, correct results will not be guaranteed. This
represents a typical shortcoming of multiple scales analysis.

Now we demonstrate how the unexpected time scales such as
$\sigma=\epsilon^{3/2} t$ appear automatically from the RG equation,
starting only with a straightforward perturbative expansion.
Substituting $a=1/4+a_1 \epsilon +a_2 \epsilon^2+\cdots$ in
\mathieu\ and expanding in powers of $\epsilon$ ({\it not}
$\epsilon^{1/2}$) as $y=y_0 +\epsilon y_1 +\epsilon^2 y_2+\cdots$, we
get
$$\frac{d^2 y_0}{d t^2} + \frac{1}{4} y_0 = 0,\eqn\muone$$
$$\frac{d^2 y_1}{d t^2} + \frac{1}{4} y_1 = -(a_1 + 2\cos t)y_0,\eqn\mutwo$$
$$\frac{d^2 y_2}{d t^2} + \frac{1}{4} y_2 = -a_2 y_0 -(a_1 + 2\cos t)y_1,
\eqn\muthree$$
and so on. First, let us determine the first order coefficient $a_1$.
The straightforward perturbation result, to $O(\epsilon)$, is given by
$$\eqalign{
y(t)=& R_0 \cos(t/2 + \Theta_0) +\epsilon R_0 \Big\{-\frac{1}{2}
\cos(t/2+\Theta_0)+ \frac{1}{2} \cos(3t/2+\Theta_0)\cr
& -a_1 (t-t_0)\sin(t/2+\Theta_0)
-(t-t_0)\sin(t/2-\Theta_0)\Big\}+O(\epsilon^2),\cr}\eqn\mufour$$
where $R_0,\Theta_0$ are constants dependent on initial conditions
given at some arbitrary time $t_0$.  Similarly, the secular divergences
can be removed by regarding $t_0$ as a regularization parameter and
renormalizing the bare amplitude $A_0$ and bare phase $\Theta_0$:
$R_0(t_0)=Z_1(t_0,\mu)R(\mu), \Theta_0(t_0)=Z_2(t_0,\mu)+\Theta(\mu)$,
where $\mu$ is some arbitrary time scale, as was done in previous
problems.  The renormalized perturbation result is
$$\eqalign{
y(t)=& \Big\{R(\mu)+\epsilon R \Big(-1/2 + (t-\mu)\sin{2\Theta(\mu)}
\Big)\Big\}\cos(t/2+\Theta)-\epsilon R\cr & \times (a_1 +
\cos{2\Theta})(t-\mu) \sin(t/2+\Theta)+\epsilon
\frac{R}{2}\cos(3t/2+\Theta)+O(\epsilon^2).\cr}
\eqn\mufive$$
The RG equation $\pd y/{\pd \mu}=0$ for any $t$
gives
$$\frac{d R}{d \mu}=\epsilon R\sin{2\Theta}+O(\epsilon^2), \quad
\frac{d \Theta} {d \mu}=\epsilon(a_1 +\cos{2\Theta})+O(\epsilon^2).
\eqn\musix$$ For convenience, we introduce the complex amplitude $A=R
e^{i\Theta}$ as $A=B + i C$, with its real and imaginary parts
$B=R\cos{\Theta}, C=R\sin{\Theta}$.  The equations for $B(\mu)$ and
$C(\mu)$ are $$B'(\mu)=\epsilon(1-a_1)C(\mu), \quad C'(\mu)=\epsilon
(1+a_1)B(\mu).  \eqn\museven$$
Thus, we have
$$B''(\mu)=\epsilon^2 (1-a_{1}^2)B(\mu).\eqn\mueight$$
Solving this and setting $\mu=t$, we get
$$B(t)=K_1 e^{\pm \sqrt{1-a_{1}^2}\ \epsilon t},\eqn\munine$$
where $K_1$ is a constant, and the first slow time scale
$\tau_1=\epsilon t$ has appeared automatically. Obviously, for
$|a_1|<1$, instability sets in, where the solution grows exponentially
with time $t$, while for $|a_1|>1$, the solutions are bounded and
stable. Therefore, near $\epsilon=0$, the stability boundary is $a=1/4
\pm \epsilon +O(\epsilon^2)$.

We now set $a_1=1$ and go to the second order to determine $a_2$. For
order $\epsilon^2$, a special solution to \muthree\ is obtained
$$\eqalign{
y_2(t)=& -R_0 \Big(a_2 -\frac{1}{2}\cos{2\Theta_0}\Big)(t-t_0)
\sin(t/2+\Theta_0)\cr
&-\frac{1}{2} R_0\sin{2\Theta_0} (t-t_0)\cos(t/2+\Theta_0)\cr
&-\frac{1}{2} R_0 (1+\cos{2\Theta_0})(t-t_0)\sin(3t/2+
\Theta_0)+\frac{1}{2} R_0 \sin{2\Theta_0}\cr
& \times (t-t_0)\cos(3t/2+\Theta_0)-\frac{3}{4}R_0(1+
\cos{2\Theta_0})\cos(3t/2+\Theta_0)\cr
& -\frac{3}{4}R_0\sin{2\Theta_0})\sin(3t/2+\Theta_0)
+\frac{1}{12}R_0 \cos(5t/2+\Theta_0).\cr}\eqn\muten$$
Extending the renormalization procedure to the second order, we find
all the secular divergences to this order can be removed completely, a
sign of the consistency of perturbative renormalizability.  Keeping
only the two lowest harmonics with prime frequency and omitting other
higher frequency terms which are not important for determining the
stability boundary, we obtain the renormalized perturbation result,
to order $\epsilon^2$,
$$\eqalign{
y(t)=& \Big\{R(\mu)+\epsilon R \Big(-1/2 +(t-\mu)\sin{2\Theta(\mu)}
\Big)-\epsilon^2 \frac{R}{2}(t-\mu)\sin{2\Theta} \Big\}\cr
& \times \cos(t/2+\Theta)+ \Big\{-\epsilon R (1+\cos{2\Theta})(t-\mu)+
\epsilon^2 R(a_2-\frac{1}{2}\cos{2\Theta})\cr
& \times (t-\mu)\Big\}\sin(t/2+\Theta)+H.F.T.,\cr}\eqn\mueleven$$
where H.F.T. represents all higher frequency terms.  The RG equation to
order $\epsilon^2$ now reads
$$\frac{d R}{d \mu}=\epsilon R\sin{2\Theta}+O(\epsilon^3), \quad
\frac{d \Theta} {d \mu}=\epsilon(1 +\cos{2\Theta})+\frac{\epsilon^2
(a_2+1/2)}{1-\epsilon/2}+ O(\epsilon^3).\eqn\mutwelve$$
Accordingly, the equations for $B(\mu)$ and $C(\mu)$ become
$$B'(\mu)=-\frac{\epsilon^2 (a_2+1/2)}{1-\epsilon/2} C(\mu),\quad
  C'(\mu)=\left[2 \epsilon +\frac{\epsilon^2
  (a_2+1/2)}{1-\epsilon/2}\right]B(\mu).  \eqn\muthirteen$$
Thus, we get
$$B''(\mu)=-\left[\frac{2 \epsilon^3
(a_2+1/2)}{1-\epsilon/2}+\frac{\epsilon^4
(a_2+1/2)^2}{(1-\epsilon/2)^2}\right]B(\mu).\eqn\mufourteen$$
Keeping only the lowest order term of \mufourteen\ gives
$$B''(\mu)\approx - \epsilon^3 (2a_2+1) B(\mu)+
O(\epsilon^4),\eqn\mufifteen$$
which has the solution (setting $\mu=t$)
$$B(t)=K_2 e^{\pm \sqrt{2a_2+1}\ \epsilon^{3/2}t},\eqn\musixteen$$
where $K_2$ is a constant, and the second and the third slow time
scales $\sigma=\epsilon^{3/2}t,\tau_2=\epsilon^2 t$ appear naturally.
We apparently have stable solutions for $a_2>-1/2$ and unstable
solutions for $a_2<-1/2$. Therefore, to order $\epsilon^2$, the
instability boundary is given by
$$a(\epsilon)=\frac{1}{4}+\epsilon-\frac{1}{2}\epsilon^2+O(\epsilon^3),\quad
\epsilon\rightarrow 0.\eqn\museventeen$$

\medskip

{\bf\section{Oscillator with Time Dependent Spring Constant}}
\smallskip

The third illustrative example is an oscillator governed by the
equation\Ref\bendertimedep{See ref. \bender, page 567.}
$$\frac{d^2 y}{d t^2}+y-\epsilon t y = 0.\eqn\bsone$$
The initial conditions are $y(0)=1$ and $y'(0)=0$.  The regular
perturbation theory breaks down for $t\rightarrow \infty$, and multiple scales analysis can be
applied to eliminate the secular behavior. However, it turns out that
multiple time scales must be chosen as $\tau_0 = t$,
$\tau_1=\epsilon^{1/2}t$, $\tau_2=\epsilon t,\cdots$. Since the
frequency of the oscillator is found to be time-dependent, the method of
stretched coordinates or the so-called method of uniformization or
renormalization (in the conventional applied mathematics sense) does not
work here.

We will see how a uniformly valid solution can be constructed simply
from the naive perturbation series with the aid of the RG.  To
solve \bsone, we assume a straightforward expansion in powers of
$\epsilon$ ({\it not} $\epsilon^{1/2}$), $y=y_0+\epsilon y_1+\epsilon^2
y_2+\cdots$. The bare perturbation result, to order $\epsilon$, is given
by
$$
y(t)=R_0 \cos(t+\Theta_0)+\epsilon R_0 \Big\{
\frac{1}{4}(t^2 - t_{0}^2)+\frac{1}{4}
(t - t_{0})\Big\}\sin(t+\Theta_0)+O(\epsilon^2).\eqn\bstwo$$
As in the preceding examples, renormalizing the bare amplitude $R_0$ and
phase $\Theta_0$ removes the secular divergences.  The renormalized
perturbation result is
$$y(t)=\Big\{R+\frac{1}{4}\epsilon R (t-\mu+a_1)\Big\}\cos(t+\Theta)
+\frac{1}{4}\epsilon R (t^2-\mu^2+b_1)\sin(t+\Theta)+O(\epsilon^2),
\eqn\bsthree$$
where $R, \Theta$ are functions of arbitrary time scale $\mu$, and
$a_1, b_1$ are arbitrary constants.  The RG equation reads
$$\frac{d R}{d \mu}=\frac{1}{4}\epsilon R + O(\epsilon^2),\quad \frac{d
\Theta} {d \mu}=-\frac{1}{2}\epsilon \mu +O(\epsilon^2).\eqn\bsfour$$
Solving \bsfour\ and setting $\mu=t$ in \bsthree\  give
$$R(t)=R(0)e^{\frac{1}{4}\epsilon t}+O(\epsilon^2 t),\quad
\Theta(t)=-\frac{1} {4}\epsilon t^2 +\Theta(0)+O(\epsilon^2
t).\eqn\bsfive$$
Thus, we obtain the uniformly valid result
$$y(t)=R(t)\cos(t+\Theta(t))+\frac{1}{4}\epsilon R(t)\left(a_1
\cos(t+\Theta) +b_1 \sin(t+\Theta)\right)+O(\epsilon^2).\eqn\bssix$$
Imposing the boundary conditions $y(0)=1,y'(0)=0$ gives $R(0)=1,
\Theta(0)=0, a_1=0, b_1 = -1$. Therefore, the final result is
$$y(t)=e^{\frac{1}{4}\epsilon t}\cos(t-\frac{1}{4}\epsilon
t^2)-\frac{1}{4} \epsilon e^{\frac{1}{4}\epsilon
t}\sin(t-\frac{1}{4}\epsilon t^2)+O(\epsilon^2), \eqn\bsseven$$
where the frequency defined as $\omega=d\Theta/{d t}$ becomes
time-dependent: $\omega=1-\frac{1}{2}\epsilon t+O(\epsilon^2)$.
Rewriting $\epsilon t^2$ as $(\epsilon^{1/2} t)^2$, two slow time
scales $T_1=\epsilon^{1/2}t$, $T_2=\epsilon t$ are easily identified from
the RG result (but these identifications are unnecessary in our
approach).

The RG scheme given above is also applicable to quantum systems with
discrete or continuous energy spectrums, especially those which involve
resonance phenomena, \eg, the Rabi flopping, the Stark shift, the
Bloch-Siegert shift\rlap.\Ref\stark{For example, see L. Allen and J. H.
Eberly, {\sl Optical Resonance and Two-level Atoms} (Wiley, New York,
N.Y., 1975); P. Meystre and M. Saregent III, {\sl Elements of Quantum
Optics} (Springer-Verlag, New York, N.Y., 1990).}  The multiple time
scale perturbation analysis has successfully given a unified framework
for all quantum resonance\rlap.\Ref\frasca{M. Frasca, \journal Il Nuovo
Cimento &107&915(92).}  In a similar way, the RG method simply recovers
all resonance equations which turn out to be simply RG equations. The
application of RG to the time-dependent Schr\"{o}dinger equation also
reproduces the Fermi's Golden Rule\rlap.\Ref\landau{For example, see L.
D. Landau and E. M.  Lifshitz, {\sl Quantum Mechanics} (Pergamon Press,
New York, 1985).}  Here we will not give detailed calculations of these
problems.  In the next section, we will show that WKB problems can be
easily solved using the RG method. Therefore, many quantum problems
which are usually solved using WKB and/or multiple scales analysis can
also be studied using the RG approach.

To summarize, it seems that the RG method is more efficient and
mechanical than the multiple scales method in determining the multiple
slow time scales. In the RG approach, the starting point is simply a
straightforward naive perturbation series, and all necessary multiple
scales arise naturally from RG equations.  The above examples reveal two
important points, demonstrated more generally below: (1) the results of
multiple scales analysis can be obtained from renormalized perturbation
theory, and (2) the RG equation describes the long time scale motion of
the amplitude and the phase.

\medskip

{\bf\chapter{Boundary-Layer Theory, WKB and RG}}
\smallskip

Another important class of singular problems is that for which the
highest order derivative of the equation is multiplied by a small
parameter $\epsilon$, \eg, WKB and boundary layer problems.

Boundary-layer theory and asymptotic matching are a collection of
singular perturbation methods for constructing a uniformly and globally
valid solution by calculating the separated outer and inner solutions
and then matching them across intermediate scale solutions.  Quite
often, the intermediate matching is very lengthy and only some
particular matching method will work.  WKB theory is well known to be a
powerful tool for obtaining a global approximation to solutions of a
linear differential equation whose highest derivative is multiplied by a
small parameter $\epsilon$. Many linear problems often solved by WKB
theory can be solved by boundary layer theory; indeed, in these cases,
boundary layer theory (thickness of the boundary layer goes to zero as
$\epsilon\rightarrow 0$) is a special case of WKB (thickness of the
boundary layer remains finite even as $\epsilon\rightarrow 0$).  The
limitation of the conventional WKB method is that it applies only to
linear problems, while boundary layer theory works for linear as well as
nonlinear problems.

In this section we will demonstrate explicitly that many boundary layer
problems, linear or nonlinear, can be solved by the RG.  The uniformly
valid asymptotics of boundary layer problems can actually be constructed
from the inner expansion alone, with the aid of the RG, without the need
for intermediate matching.
\medskip

{\bf\section{Simple Linear Example}}
\smallskip

Consider the following simple example, which describes the
motion of an overdamped linear oscillator:
$$\epsilon \frac{d^2 y}{d t^2} + \frac{d y}{d t} + y =0, \qquad
\epsilon \ll 1,
\eqn\one$$
where $\epsilon$ is a small parameter.  A standard dominant-balance
argument shows that there exists a boundary layer of thickness
$\delta=O(\epsilon)$ at $t=0$.  Thus, we set $t=\epsilon \tau$, and
rewrite equation \one\ as
$$\frac{d^2 y}{d \tau^2} + \frac{d y}{d \tau} + \epsilon y=0.
\eqn\five$$
Naive expansion gives
$$y(\tau)=A_0+B_0 e^{-\tau}+\epsilon \big [-A_0(\tau-\tau_0)+B_0
(\tau-\tau_0) e^{-\tau}\big ]+O(\epsilon),
\eqn\six$$
where the coefficients $A_0,B_0$ are constants of integration and
$O(\epsilon)$ refers to all the regular terms of order $\epsilon$ and
higher, which are finite even in the limit $\tau-\tau_0 \rightarrow \infty$.
This naive perturbation
theory breaks down due to the divergence of secular terms for
large $\tau-\tau_0$. However, this divergence can be removed by
regarding $\tau_0$ as a regularization parameter and renormalizing
$A_0,B_0$ as $A_0(\tau_0)=Z_1 A(\mu)$, and $B_0(\tau_0)=Z_2 B(\mu)$.
Here $\mu$ is an arbitrary time, and $A,B$ are the renormalized
counterparts of $A_0,B_0$. The renormalization constants
$Z_1=\sum_{0}^{\infty} a_n(\tau_0,\mu) \epsilon^{n},
Z_2=\sum_{0}^{\infty} b_n(\tau_0,\mu)\epsilon^{n}$ ($a_0=1,b_0=1$) are
chosen order by order in $\epsilon$ to eliminate the secular
divergences.  Split $\tau-\tau_0$ as $(\tau-\mu)+(\mu-\tau_0)$, and then
absorb the divergent part $\mu-\tau_0$ in the limit $\tau_0 \rightarrow
-\infty$ by redefining $A_0$ and $B_0$.  Choosing $a_1=\mu-\tau_0,
b_1=-(\mu-\tau_0)$, we get the renormalized perturbation result
$$y(\tau)= A(\mu)-\epsilon A(\mu)(\tau-\mu)+\big [ B(\mu)+\epsilon
B(\mu) (\tau-\mu)\big ] e^{-\tau}+O(\epsilon).\eqn\eight$$ However, it
is impossible that the actual solution $y(\tau)$ can depend on the
arbitrary time $\mu$ which is not present in the original problem.  Thus
we have the renormalization group equation $\pd y/\pd \mu =0$ for any
$\tau$, which gives
$$\frac{d A}{d \mu}+\epsilon A
+\left[\frac{d B}{d \mu}-\epsilon B \right]
e^{-\tau}+O(\epsilon^2)=0,\eqn\nine$$
or
$$\frac{d A}{d \mu}=-\epsilon A+O(\epsilon^2),\quad \frac{d B}{d \mu}
=\epsilon B+O(\epsilon^2).\eqn\ten$$
Extending the RG calculation to the second order gives, without
any ambiguity,
$$\frac{d A}{d \mu}=-(\epsilon A+\epsilon^2 A)+O(\epsilon^3),\quad
\frac{d B}{d \mu}=\epsilon B+\epsilon^2 B +O(\epsilon^3).\eqn\tencomp$$
Solving them, setting $\mu=\tau$ and setting back $\tau=t/\epsilon$ in
\eight, we finally obtain the uniformly valid solution
$$y(t)=C_1 e^{-(1+\epsilon)t}+C_2 e^{-t/\epsilon+(1+\epsilon)t}
+O(\epsilon^2),\eqn\eleven$$
where $C_1,C_2$ are constants to be determined by the initial
conditions.  Clearly, the RG result to order $\epsilon^2$ recovers
exactly that obtained by the standard singular
methods\rlap.\refmark{\bender} Notice that the equations in \tencomp\
are nothing but the equations of motion for slow time scale: the
amplitude equations. Thus, amplitude equations are renormalization group
equations.  We announced this result previously, and derived the Burgers
equation as a renormalization group equation.\refmark{\kawasakifest}  A
much more complicated example illustrating this point will be given in
Section 5.

\medskip

{\bf\section{Example with log $\bmath\epsilon$}}
\smallskip

The second example we consider is\Ref\benderlog{See ref. \bender, page
442.}
$$\epsilon y''+ xy'-xy=0, \quad y(0)=0, y(1)=e. \eqn\logone$$
A standard dominant-balance argument tells us that there exists a
boundary layer of thickness of order $\epsilon^{1/2}$ (but {\it not\/}
$\epsilon$) at $x=0$. The complication in the conventional asymptotic
matching stems from the fact that the inner expansion must contain not
only powers of $\epsilon^{1/2}$ but also those terms containing
combinations of $\epsilon$ and $\log \epsilon$ to make the intermediate
matching successful. Here we explicitly show that the renormalized naive
inner expansion in powers of $\epsilon^{1/2}$ gives a uniformly valid
asymptotic solution.  This reveals that those unexpected terms
containing $\log \epsilon$ in the conventional approach are just an
artifact of perturbative expansions of $x^{-\epsilon}$.

Assuming $x=\epsilon^{1/2} X$, and $y(x)=Y(X)$, we transform
\logone\ into
$$\frac{d^2 Y}{d X^2} + X \frac{d Y}{d X} - \sqrt{\epsilon} X Y=0, \quad
Y(0)=0, Y(1/\sqrt{\epsilon})=e. \eqn\logtwo$$
Naive expansion in $\epsilon^{1/2}$, $Y(X)=Y_0(X)+
\epsilon^{1/2}Y_1(X)+\epsilon Y_2(X)+\cdots$ gives
$$ Y_0''+XY_0'=0,\quad Y_n''+XY_n'=XY_{n-1},\quad (n\ge
1),\eqn\logthree$$
Thus the naive perturbation result to order $\epsilon$ is
$$\eqalign{ Y(X)\sim & A_0 +B_0 \int_{0}^{X} ds e^{-s^2/2} +
\epsilon^{1/2}\Big\{ A_0 (X-X_0)+B_0 (X-X_0)\int_{0}^{X} ds
e^{-s^2/2}\cr &+ R.T.\Big\}+\epsilon \Big\{\frac{1}{2}A_0 (X-X_0)^2 +
\frac{1}{2} B_0 (X-X_0) ^2 \int_{0}^{X} ds e^{-s^2/2}\cr
&-\left(\frac{2}{\sqrt{\pi}}A_0+B_0\right)\log\left(\frac{X}{X_0}\right)
\int_{0}^{X} ds e^{-s^2/2}+R.T.\Big\},\cr}\eqn\logfour$$
where $A_0, B_0$ are integration constants, and $R.T.$ represents
regular terms finite even in the limit $X-X_0\rightarrow \infty$ and
$\log(X/X_0) \rightarrow \infty$. The divergence can be controlled by
renormalizing $A_0=Z_1 A(\mu)$, $B_0=Z_2 B(\mu)$, where
$Z_1(\mu)=\sum_0^{\infty} a_n \epsilon^{n/2}, a_0=1$ and $Z_2(\mu)=
\sum_0^{\infty} b_n \epsilon^{n/2}, b_0=1$, are renormalization
constants and  $\mu$ is some arbitrary position. The choice
$a_1=X_0-\mu, a_2=(1/2)(X_0-\mu)^2$ and $b_1=X_0-\mu,
b_2=(1/2)(X_0-\mu)^2- (\frac{2}{\sqrt{\pi}}A+B)\log(X_0/\mu)$
successfully removes the divergences up to order $\epsilon$, and the
renormalized perturbation result is
$$\eqalign{
Y(X)\sim & \Big\{A(\mu)+\epsilon^{1/2}A(X-\mu)+\epsilon
\frac{1}{2}A(X-\mu)^2 \Big\}+\Big\{B(\mu)+\epsilon^{1/2}B(X-\mu)\cr
& +\epsilon \frac{1}{2}B(X-\mu)^2 -\epsilon
(\frac{2}{\sqrt{\pi}}A+B)\log(X/\mu)\Big\}\int_{0}^{X} ds e^{-s^2/2}
.\cr}\eqn\logfive$$
The RG equation ${\pd Y}/{\pd \mu}=0$ gives
$$\frac{d A}{d \mu}=\epsilon^{1/2} A + O(\epsilon^{3/2}),\eqn\logsix$$
$$\frac{d B}{d \mu}=\epsilon^{1/2} B -\epsilon (\frac{2}{\sqrt{\pi}}A+B)/\mu
+ O(\epsilon^{3/2}).\eqn\logseven$$
Solving these two equations, we obtain
$$A(\mu)=C_1 e^{\epsilon^{1/2}\mu}+O(\epsilon^{3/2}\mu),\eqn\logeight$$
$$B(\mu)=-\frac{\epsilon}{1+\epsilon}\frac{2}{\sqrt{\pi}}C_1 \mu
e^{\epsilon^{1/2}\mu}+C_2 \mu^{-\epsilon} e^{\epsilon^{1/2}\mu}
+O(\epsilon^{3/2}\mu),\eqn\lognine$$
where $C_1,C_2$ are constants to be determined by the given boundary
conditions. Setting $\mu=X$, we obtain
$$Y(X)\sim C_1 e^{\epsilon^{1/2}X}+\Big\{-\frac{\epsilon}{1+\epsilon}\frac{2}
{\sqrt{\pi}}C_1 X +C_2 X^{-\epsilon}\Big\} e^{\epsilon^{1/2}X}\int_{0}^{X}
ds e^{-s^2/2},\eqn\logten$$
Imposing boundary conditions $Y(0)=0, Y(1/\sqrt{\epsilon})=e$ gives $C_1=0$
and $C_2=\sqrt{2/\pi}\epsilon^{-\epsilon/2}=\sqrt{2/\pi}e^{-\frac{1}{2}
 \epsilon \log\epsilon}$ as $\epsilon\rightarrow 0_{+}$.
Setting back $X=x/\epsilon^{1/2}$, we obtain the final uniformly valid
asymptotic result to order $\epsilon$,
$$y(x)\sim e^{x} x^{-\epsilon}\Big\{1-\sqrt{2/\pi}
\int_{x/\sqrt{\epsilon}}^{\infty} ds e^{-s^2/2} \Big\}.\eqn\logeleven$$
Thus the terms such as $\epsilon \log\epsilon$, which are present in
the inner expansion given in (\eg,) ref. \bender, are relics of the
expansion of $x^{-\epsilon}$. It is worthwhile to note that the RG
result is slightly different from the asymptotic matching result given
by Bender and Orszag in their book\rlap.\refmark{\benderlog} To
leading order, the former is
$$y_{0}^{RG}(x)\sim e^{x} \Big\{1-\sqrt{2/\pi}
\int_{x/\sqrt{\epsilon}}^{\infty} ds e^{-s^2/2} \Big\},\eqn\logtwelve$$
while the latter is
$$y_{0}^{BO}(x)\sim e^{x} -\sqrt{2/\pi}
\int_{x/\sqrt{\epsilon}}^{\infty} ds e^{-s^2/2}.\eqn\logthirteen$$
Comparing with the numerical result of the original equation \logone, we
find that in the boundary-layer region, the RG result \logtwelve\ is a
better approximant than the standard result \logthirteen.

\medskip

{\bf\section{Nonlinear Boundary Layer Problem}}
\smallskip

Boundary-layer analysis applies to nonlinear as well as to linear
differential equations. In this section and in the following section,
we will demonstrate that the RG method can be used to solve nonlinear 
boundary layer problems. 

Let us consider the following
illustrative nonlinear problem\rlap:\Ref\bendernonlin{See ref. \bender,
page 463.}
$$\epsilon y''+2y'+e^{y}=0,\quad y(0)=y(1)=0.\eqn\nlone$$
There is only one boundary layer of thickness $\epsilon$ at $x=0$. Setting
$X=x/\epsilon, Y(X)=y(x)$ in \nlone\ gives
$$\frac{d^2 Y}{d X^2}+2\frac{d Y}{d X}=-\epsilon e^{Y}.\eqn\nltwo$$
Assuming an inner expansion $Y=Y_0+\epsilon Y_1+\cdots$ gives the following
asymptotic result as $X\rightarrow \infty$,
$$Y(X)\sim A_0+B_0 e^{-2X}-\epsilon
\left\{\frac{1}{2}e^{A_0}(X-X_0)+R.T.\right\}
+O(\epsilon^2),\eqn\nlthree$$
where $A_0, B_0$ are integration constants and $R.T.$ represents all
regular terms in the expansion finite even in the limit
$X-X_0\rightarrow \infty$.  The renormalized perturbation result
obtained as in the previous examples is
$$Y(X)\sim A(\mu)+B(\mu) e^{-2X}-\epsilon
\frac{1}{2}e^{A(\mu)}(X-\mu)+O(\epsilon^2).\eqn\nlfour$$
The RG equation gives, to order $\epsilon$,
$$\frac{d A}{d \mu}+\epsilon \frac{1}{2}e^{A}=0,\quad \frac{d B}{d \mu}=0.
\eqn\nlfive$$
Solving \nlfive\ , we get
$$A(\mu)=\log\left({\frac{2}{\epsilon \mu +C_1}}\right),
\quad B(\mu)=C_2, \eqn\nlsix$$
where $C_1, C_2$ are constants of integration to be determined by the
given boundary conditions.  Equating $\mu$ and $X$ in \nlfour\  and
restoring $x=\epsilon X$, we obtain the uniformly valid asymptotic
result
$$y(x)\sim \log\left({\frac{2}{x+C_1}}\right)+C_2
e^{-2x/\epsilon}+O(\epsilon).  \eqn\nlseven$$
Imposing boundary conditions $y(0)=0, y(1)=0$ gives $C_1=1,
C_2=-\log{2}$ in the limit $\epsilon\rightarrow 0_{+}$. Therefore, the
final result is
$$y(x)\sim \log\left({\frac{2}{x+1}}\right)-(\log{2}) e^{-2x/\epsilon}
+O(\epsilon). \eqn\nleight$$
This RG result recovers the leading order result from boundary layer
analysis.

\medskip

{\bf\section{Nonlinear Problem of Carrier}}
\smallskip

In this section, we consider a first-order nonlinear model problem of 
Carrier\rlap,\Ref\carrier{See ref. \hinch, page 103.}
$$(x+\epsilon f)f' + f =1, \qquad\qquad f(1)=2, \quad 0\leq x\leq 1.
\eqn\crone$$
The exact solution can be obtained by integrating \crone\ once, 
$$f(x,\epsilon)= -\frac{x}{\epsilon} + \left(\frac{x^2}{\epsilon^2}
+\frac{2(x+1)}{\epsilon}+4\right)^{1/2}.\eqn\crtwo$$
It becomes, however, a nontrivial singular perturbation problem, if we
pretend that we cannot obtain the exact solution.  The method of
strained coordinates or the method of asymptotic matching can be applied
with a rather lengthy matching.  We show here how to solve the problem
using RG without matching, and give the exact result, starting only from
the inner expansion.

First, we apply the usual dominant balance argument to make the
structure of the equation clear.  We introduce $X \equiv \eta(\epsilon)
x$ and $F = \delta(\epsilon) f$; the latter is needed because the
equation is nonlinear.  The original equation reads $ (X + (\epsilon
\eta/\delta) F) dF/ dX + F = \delta$.  $\epsilon \eta/\delta \ll 1$
corresponds to the outer limit, and $\epsilon \eta/\delta \sim 1$ is the
only nontrivial alternative possibility.  Hence, $\delta \sim
\epsilon^{\alpha}$ and $\eta \sim \epsilon^{\alpha -1}$ with $\alpha \in
(0,1]$ are the useful scalings.  The expansion parameter becomes $\delta
\sim \epsilon^{\alpha}$.  It turns out that any choice of $\alpha$ is
admissible in this case, and so we adopt the simplest choice $\alpha =
1$.
 
Accordingly, we rescale $f$ as $F = \epsilon f$ to convert the original
equation \crone\ to 
$$ (x+F)F' + F =\epsilon.\eqn\crthree$$
Expanding $F$ as $F=F_0 + \epsilon F_1 + \cdots$, we have
$$(x+F_0)F'_{0} + F_0 = 0, \eqn\crfour$$
whose general positive solution is 
$$F_0(x)= (x^2 + A_0)^{1/2}-x, \eqn\crfive$$
with $A_0$ a constant of integration determined by the initial
condition given at some arbitrary $x_0$. 
The first order equation is given by
$$(x+F_0)F'_{1}+ F_1 F'_{0} + F-1 = 1.\eqn\crsix$$
This linear equation has a general solution
$$F_1(x)=\frac{x-x_0}{(x^2+A_0)^{1/2}}.\eqn\crseven$$ 
Thus, the straightforward perturbation result, to $O(\epsilon)$, is
given by 
$$F(x)= (x^2 + A_0)^{1/2}-x + \epsilon \frac{(x-x_0)}{(x^2+A_0)^{1/2}} + 
O(\epsilon^2). \eqn\creight$$
We see that this naive perturbation \creight\ breaks down formally for
$x \gg x_0 $.  Actually, the domain of our problem is finite, and
because $x$ is not scaled, it is not possible that $x \gg x_0$ can occur
within the domain.  A better argument is as follows. Since the boundary
condition is $F(1) = 2\epsilon$, near $x =1$ the $O(\epsilon)$ term
dominates; this is a singular perturbation, and indeed the perturbation
term diverges relative to the zeroth order term.

The secular divergence can be removed 
by renormalizing $A_0$ by $A_0(x_0)=Z A(\mu)$, and the renormalized
perturbation result obtained is
$$ F(x)= (x^2 + A(\mu))^{1/2}-x + \epsilon 
\frac{(x-\mu)}{(x^2+A(\mu))^{1/2}} + O(\epsilon^2).\eqn\crnine$$
The RG equation gives, to $O(\epsilon)$,
$$dA/d\mu = 2 \epsilon\eqn\creleven$$
with solution 
$$A(\mu)=A(0)+ 2 \epsilon \mu. \eqn\crtwelve$$
Setting $\mu=x$ and $f=F/\epsilon$, we obtain the uniformly valid
asymptotic result 
$$   
f(x,\epsilon)= -\frac{x}{\epsilon} + \left(\frac{x^2}{\epsilon^2}
+\frac{2
x}{\epsilon}+\frac{A(0)}{\epsilon^2}\right)^{1/2}. \eqn\crthirteen$$
Imposing the boundary condition $f(1)=2$ gives $A(0)= 2 \epsilon +4
\epsilon^2$. Therefore, the uniformly valid result to order
$\epsilon$, is given by
$$ f(x,\epsilon)= -\frac{x}{\epsilon} + \left(\frac{x^2}{\epsilon^2}
+\frac{2(x+1)}{\epsilon}+4\right)^{1/2}. \eqn\crfifteen$$
This happens to be the exact solution to the problem.
A further calculation demonstrates that all the higher order
corrections vanish. The conventional methods can also recover the exact
result, but clearly the RG is simpler.

{\bf\section{Problem with Multiple Boundary Layers}}
\smallskip

In many situations, there exist multiple boundary layers at one side,
for which multiple calculations of inner and outer solutions and their
asymptotic matchings have to be made in different separated regions to
obtain a uniformly valid solution. Again it turns out that the RG
method manages to produce the solution without any matching needed. Let
us consider the following initial-value problem\Ref\murtwo{See
ref. \murdock, page 388, example 7.6.1.}
$$\epsilon^{3/2} y'''+(\epsilon^{1/2}+\epsilon+\epsilon^{3/2})y''+
(1+\epsilon^{1/2}+\epsilon)y'+y=0,\eqn\doublelayer$$
with initial conditions $y(0)=3, y'(0)=-1-\epsilon^{-1/2}-
\epsilon^{-1}, y''(0)=1+\epsilon^{-1}+\epsilon^{-2}$.\refmark{\bender}
The exact solution is
$y(t)=e^{-x}+e^{-x/\epsilon^{1/2}}+e^{-x/\epsilon}$.  Pretending we do
not know how to solve it exactly, we resort to conventional singular
perturbation methods. It turns out that the conventional perturbation
calculation is very tedious and rather challenging. By dominant balance,
this problem is found to have two distinguished boundary layers at
$t=0$, of thickness of order $\epsilon^{1/2}$ and $\epsilon$,
respectively. Therefore, one outer solution and two inner solutions must
be calculated and two asymptotic matchings are necessary, if
boundary layer theory is used. Starting only with the thinnest
or innermost
boundary layer by rescaling $t$ by $t=\epsilon T$, and
expanding $y=Y(T)$ ({\it e.g.},) in $\epsilon^{1/2}$, the RG method
successfully recovers the exact solution without any matching.

\medskip

{\bf\section{Linear Boundary-layer and WKB problems I: no turning points}}
\smallskip

To conclude this section, we show how linear boundary layer and WKB
problems in general forms can be treated using RG in a unified fashion.
This relationship between boundary layer theory and WKB is explained in
Bender and Orszag's book\rlap.\refmark{\bender} The boundary-layer type
problem we wish to study using RG has the following general form:
$$\epsilon^2 \frac{d^2 y}{d x^2}+ a(x)\frac{d y}{d x}-b(x) y=0, \quad
0\le x \le 1, \quad \epsilon \rightarrow 0_{+}, \eqn\lone$$ where we
assume that $a(x)$ is differentiable and $b(x)$ is an arbitrary, not
necessarily continuous function.  This equation covers all linear
examples we presented earlier in this section.  A simple
dominant-balance argument determines that in general, the boundary layer
lies at $x=0$ when $a(x)\ge 0$ for $0\le x \le 1$, and that the boundary
layer lies at $x=1$ when $a(x)< 0$ for $0\le x \le 1$.  Without loss of
generality we will consider only the former case.

Although in a number of cases we could perform perturbative RG analysis
on the original general equation \lone, often it is wiser to start with
the canonical form of equation \lone\ under the transformation
$$y(x)=\exp \left[-\frac{1}{2\epsilon^2}\int^{x} a(x')dx'\right] u(x),
\eqn\ltwo$$
converting \lone\ to
$$\epsilon^2 \frac{d^2 u}{d x^2}= Q(x) u(x), \eqn\lthree$$
with
$$Q(x)\equiv \frac{1}{4\epsilon^2}a^2(x)+\frac{1}{2}a'(x)+b(x). \eqn\lfour$$
This form is just the Schr\"odinger form, which can be solved by WKB
methods.  Consequently, we can treat both linear boundary layer and WKB
problems in a unified way.

In the remainder of this section and in the following section, we will
show how to solve Schr\"odinger equations using RG.  Our strategy is to
first introduce a natural change of the independent variable which
allows one to obtain efficiently the non-perturbative part of the
solution.  The transformation is identical to the independent variable
portion of the standard Liouville-Green transformation\Ref\liouville{J.
Liouville \journal J. Math. Pure Appl. &2&16(1837); G. Green, \journal
Trans. Camb. Phil. Soc. &6&457(1837).} or its natural generalization
used by Langer\rlap,\Ref\langerre{R.E. Langer, \journal Trans. Am.
Math.  Soc., &33&23(31); \journal ibid &36&90(34); \journal ibid
&37&397(35).}  but the crucial difference is that we do not introduce
the new dependent variable. This is the analogue of the
geometrical-optics approximation in WKB theory\rlap,\refmark{\bender}
and is the starting point of a renormalized perturbation series, which
reproduces the physical-optics and higher-order WKB approximations.
Although it may be possible to derive even the geometric-optics
approximation using RG, we have not succeeded in so doing.  
transformation is sensitive 
function $Q(x)$.  When $Q(x)$ vanishes, its zeros lead to turning points
in the standard WKB approach.  The simplest WKB approximations break
down there, and connection formulae are required in the conventional
procedure in order to match approximations on either side.  
other hand, our procedure leads to a uniformly valid approximation.
Langer\refmark{\langerre} found that a suitable generalization of the
Liouville-Green transformation can produce a uniformly valid
approximation across the turning point.  Again for the cases with
turning points, we transform the independent variable only with a
straightforward generalization of the no-turning point case.  We
emphasize that we are able to avoid the need to perform matching, and
that the transformation of the dependent variable is produced naturally
by RG.  The use of RG is {\em not} responsible for the choice of the
transformation of the independent variable, but our choice not to
introduce the transformation of dependent variables in contrast to the
approaches by Liouville and Green and Langer is motivated by RG.  This
allows us to choose a better transformation of the dependent variable,
which agrees with the conventional result in the small $\epsilon$
limit.  The corrections and prefactors which accompany the zeroth order
Langer-type solution {\it are\/} calculated by RG, and do differ from
and improve upon those obtained by the standard analysis.  Furthermore,
we can expand our asymptotic sequence in $\epsilon$ to reproduce the
standard textbook results.

The remainder of this section concerns Schr\"odinger problems with no
turning points.  The following section discusses the case with one
turning point, and gives an outline of how the methods can be
generalized for higher numbers of turning points and for
multiple-boundary-layer linear problems as well.

We will first rederive
the well-known physical-optics approximation using the RG theory,
valid when the function $Q(x)$ has no zeroes in the interval
of interest. 
Following Liouville and Green, we introduce a new independent variable 
$t=f(x)$ implicitly determined
as $dt=\sqrt{Q} dx/\epsilon$.  
The choice is natural from the perturbation
point of view, because even when $du/d(x/\epsilon)$ is
significant, $dQ(x)/dx$ is not, so $Q(x)$ can be regarded as a constant to
order $O(\epsilon^0)$.
Eq. \lthree\ is thus converted to
$$\frac{d^2 u}{d t^2} - u = 2\epsilon S(x) \frac{d u}{d t},
\eqn\lfive$$
where $S(x)\equiv -(1/4) Q^{-3/2}Q'(x)$ is assumed to be a slowly
varying function on the time scale $t$, of order unity, and
$S(x)\not=0$ for $0\le x \le 1$.

Naively expanding $u$ as $u(t)=u_0(t)+\epsilon u_1(t)+\cdots$, we get
the bare perturbation result
$$\eqalign{
u(t)=& e^{t}\Big\{A_0 + \epsilon A_0 \int_{t_0}^{t}S\left(x(t')\right)
dt'-\epsilon A_0 e^{-2t}\int_{t_0}^{t}S\left(x(t')\right) e^{2t'}dt'\Big\}\cr
& +e^{-t}\Big\{B_0 + \epsilon B_0 \int_{t_0}^{t}S\left(x(t')\right)dt'
-\epsilon B_0 e^{2t}\int_{t_0}^{t}S\left(x(t')\right) e^{-2t'}dt'\Big\}
+O(\epsilon^2),\cr}\eqn\lsix$$
where $A_0, B_0$ are integration constants.  The corresponding
renormalized result is
$$
u(t)=e^{t}\Big\{A(\mu) + \epsilon A(\mu) \int_{\mu}^{t}S dt'\Big\}
+e^{-t}\Big\{B(\mu) + \epsilon B(\mu) \int_{\mu}^{t}S
dt'\Big\}+O(\epsilon),\eqn\lseven$$
where $O(\epsilon)$ refers to all regular terms of order $\epsilon$
which remain finite even as $t-t_0 \rightarrow \infty$.  The RG
equation ${\pd u}/{\pd \mu}\equiv 0$ gives
$$ \frac{d C}{d \mu}+\epsilon \frac{1}{4}Q^{-3/2}Q'(x(\mu)) C=O(\epsilon^2),
\eqn\leight$$
where $C=A$ or $B$. Again, equation \leight\ corresponds to the
amplitude equation or slow motion equation.  Setting $\mu=t$ and using
$dt=Q^{1/2}dx/\epsilon$, we get
$$A(x)\sim Q^{-1/4}(x),\quad\quad B(x)\sim Q^{-1/4}(x).\eqn\lten$$
This is exactly the adiabatic invariant
$A(x)Q^{1/4}(x)=A(0)Q^{1/4}(0)=$ constant.  The physical-optics
approximation for WKB equation \lthree\ is recovered
$$u(x)\sim C_1 Q^{-1/4}(x) \exp\left[\frac{1}{\epsilon}\int^{x}
dx'\sqrt{Q(t')} \right]+C_2 Q^{-1/4}(x)
\exp\left[-\frac{1}{\epsilon}\int^{x} dx'\sqrt{Q(t')}
\right],\eqn\leleven$$
as $\epsilon \rightarrow 0$.

The uniformly valid asymptotic result $y(x)$ for the general linear
boundary layer problem \lone\ is given by \ltwo.  For numerical
evaluation of (3.33), we do not need any further expansion, because
\ltwo\ is the uniformly valid result we want.  To compare, however, with
the conventional results due to asymptotic matching methods, let us make
asymptotic expansions of $Q(x)$.

As a simple check, let us assume that $a(x),b(x)$ are some analytic
functions, and $a(x)>0$ for $0\le x \le 1$, and $a(0)\not=0$.
Obviously, in the whole region $0\le x \le 1$, as $\epsilon \rightarrow
0$, the term $a^{2}(x)/4\epsilon^2$ is the dominant term, compared to
$a'(x)/2$ and $b(x)$. Simply Taylor expanding as
$$\sqrt{Q(x)}\simeq \frac{a(x)}{2\epsilon} + \frac{\epsilon}{2}\frac{a'(x)}
{a(x)}+\epsilon \frac{b(x)}{a(x)},\eqn\ltwelve$$
and imposing boundary conditions $y(0)=A, y(1)=B$, we obtain
$$y(x)\sim B e^{\int_{1}^{x}\frac{b(\xi)}{a(\xi)}d\xi}+\frac{a(0)}
{a(x)}\left[A-B e^{\int_{1}^{0}\frac{b(\xi)}{a(\xi)}d\xi}\right]
e^{-\int_{0}^{x}\left[\frac{a(\xi)}{\epsilon^2}+\frac{b(\xi)}
	 {a(\xi)}\right]d\xi}.\eqn\lthirteen$$
This expression can be simplified further, because the second term
contributes appreciably only when $x=O(\epsilon^2)$
$(\epsilon\rightarrow 0)$. Thus,
$$y(x)\sim B \exp\left[\int_{1}^{x}\frac{b(\xi)}{a(\xi)}d\xi \right]+
\left[A-B
\exp\int_{1}^{0}\frac{b(\xi)}{a(\xi)}d\xi\right]e^{-a(0)x/\epsilon^2}.
\eqn\lfourteen$$
This is exactly the same as the uniformly valid leading boundary layer
or WKB result.

It is known that the case with $a(0)=0$ is subtle.  For simplicity, we
consider only the cases $a(x)=x^{\alpha}, b(x)=1$, where $\alpha > -1$
so that there exists a boundary layer at $x=0$.

When $\alpha > 1$, the thickness of boundary layer is of order
$\delta=O(\epsilon)$. When $x \gg \epsilon$, the term
$a^2(x)/4\epsilon^2$ dominates over other two terms, $a'(x)/2$ and
$b(x)$, in $Q(x)$. However, when $x\sim O(\epsilon)$, we have to be
careful with the asymptotic expansion of $Q(x)$, since the dominant
term now is $b(x)=1$. Thus, as $\epsilon \rightarrow 0$, the leading
term of $\sqrt{Q(x)}$ is $1$. The final uniformly valid approximation
is
$$y(x)\sim B \exp \left[ \frac{1}{\alpha-1}(1-x^{1-\alpha})\right]+
A \exp \left[ -x/\epsilon\right].\eqn\bltypetwo$$

When $|\alpha|\leq 1$, it is straightforward to check that the boundary
layer is of thickness of order $\delta \sim \epsilon^{2/({1+\alpha})}$,
and that the first and second term in $Q(x)$ are of the same order, when
$x \sim \delta(\epsilon)$. The uniformly valid expression turns out to
be
$$\eqalign{
y(x)=& B \exp \left[ \int_{1}^{x}\frac{b(\xi)}{a(\xi)} d\xi \right]+
\left[ A-B \exp \int_{1}^{0}\frac{b(\xi)}{a(\xi)} d\xi \right]\cr
& \times \{\frac{Q(0)}{Q(x)}\}^{-1/4}
\exp \left[-\int_{0}^{x} d\xi \{\frac{a(\xi)}{2 \epsilon^2}+
\frac{\sqrt{Q(\xi)}}{\epsilon}\}\right].\cr}
\eqn\bltypethree$$
Expanding the above leading uniformly valid result obtained with the aid
of RG recovers the outer and inner solutions due to boundary layer
theory and asymptotic matching. Note that the above results are obtained
from the ``inner expansion" alone without ever having to perform any
asymptotic matching.  This is practically important as we will see in
the next section.  However,  the main message here is conceptually more
important:  conventional singular perturbation methods can be understood
naturally as the standard renormalized perturbation procedure.

\section{WKB analysis II: turning points}

In order to complete this section, we begin by presenting a general 
discussion of Schr\"odinger equations and one-turning-point WKB problem,
and at the end of this section, we generalize the case to 
multiple-turning-point and multiple-boundary-layer problems.

The Schr\"odinger equation which we will consider in this section is
$$\epsilon^2 \frac{d^2 u}{d x^2}= Q(x) u(x), \quad u(+\infty)= 0 .
\eqn\tpone$$ When $Q$ in \tpone\ vanishes or changes its sign, the
approach in the preceding subsection fails as can easily be seen from
the presence of the factor $Q^{-1/4}$.  If $Q$ has an isolated zero at
$x = 0$ of order $\alpha >0$, we can write locally $Q(x) = x^\alpha
\psi(x)$ with a positive definite function $\psi$ without any loss of
generality. A natural choice of the counterpart of the Liouville-Green
transformation $x \rightarrow t$ is to remove the zeros from $dt/dx$:
we introduce a new independent variable $t =f(x)$ implicitly determined
as $dt = \sqrt{Q/t^\alpha} dx/\epsilon$, and integrating it gives
$$t(x)= \left(\frac{2+\alpha}{2 \epsilon} \int_{0}^{x} dx'
\sqrt{Q(x')}\right)^{2/(2+\alpha)}.\eqn\tpthree$$
The original equation \lthree\ is transformed into
$$\frac{d^2 u}{d t^2}= t^{\alpha} u + \epsilon S\left(t(x)\right)
\frac{d u}{d t}, \eqn\tpfour$$
where $S \equiv  d[(t^\alpha/Q)^{1/2}]/dx$. Since 
$t \sim x$ as $x \rightarrow 0$, $S$ is a bounded function even near 
$x=0$.
Notice that in contrast to the conventional approaches due to Liouville
and Green or Langer, we do not introduce the transformation for the
dependent variable, which will be produced by the RG procedure.  Here we
work out the simplest case $\alpha = 1$.

Expanding naively $u$ in powers of $\epsilon$ as $u=u_0+\epsilon u_1 + 
\epsilon^2 u_2 + \cdots$, we obtain the bare perturbation result to
order $\epsilon$,
$$u= C_0 \Ai(t) -\epsilon C_0 \pi \Big\{\Ai(t) \int_{t_0}^{t} dt'
S(t') \Ai'(t')\Bi(t')- \Bi(t) \int_{t_0}^{t} dt' S(t')\Ai(t') \Ai'(t')\Big\},
\eqn\tpsix$$ 
where $\Ai, \Bi$ are two linearly independent Airy functions,
and the $\Bi(t)$ function is already discarded in the zeroth order
solution, since it does not satisfy the physical condition
$u(+\infty)=0$.  In the limit $t-t_0 \rightarrow +\infty$, 
the second term of the first order perturbation $\Bi(t)
\int_{t_0}^{t} dt' S(t')\Ai(t') \Ai'(t')$ remains finite.  However, 
the term
$\int_{t_0}^{t} dt' S(t') \Ai'(t')\Bi(t')$ diverges and must be
renormalized, giving the renormalized perturbation series
$$ u= \Ai(t)\left(C(\mu)-\epsilon C(\mu)\pi \int_{\mu}^{t} dt'
S(t') \Ai'(t')\Bi(t')\right)+O(\epsilon),\eqn\tpseven$$ where $C(\mu)$ is
the counterpart of the bare amplitude $C_0(t_0)$, and $O(\epsilon)$
refers to all finite regular terms of order $\epsilon$ even in the limit
$t-t_0\rightarrow \infty$. The RG equation $du/d\mu\equiv 0$ gives
$$\frac{d C(\mu)}{d\mu} + \epsilon C(\mu)\pi S(\mu) \Ai'(\mu)
\Bi(\mu) = O(\epsilon^2). \eqn\tpeight$$
Integrating \tpeight\ and setting $\mu=t$, we get
$$C(t)=C(0)\exp\Bigg\{-\pi \int_{0}^{t} dt' \Ai'(t')\Bi(t')\frac{ d}{dt'}
\Big\{\log\left[(t'/Q)^{1/2}\right]\Big\} \Bigg\}, \eqn\tpnine$$
where $C_0$ is a constant of integration to be determined by boundary
condition at $t=0$. Thus we have arrived at the adiabatic invariant  
$$C(t)\exp\Bigg\{\pi \int_{0}^{t} dt' \Ai'(t')\Bi(t')\frac{ d}{dt'}
\Big\{\log\left[(t'/Q)^{1/2}\right]\Big\} \Bigg\}, \eqn\tpzero$$ 
which differs from that usually obtained, leading to the
the final uniformly valid solution 
$$ u = C(0)\exp\Big\{-\pi \int_{0}^{t} dt' \Ai'(t')\Bi(t')\frac{ d}{dt'}
\Big(\log\big[\left(t'/Q\right)^{1/2}\big]\Big) \Big\} \Ai(t),\eqn\tpeleven$$
where $t(x)= \left(\frac{3}{2 \epsilon} \int_{0}^{x} dx'
\sqrt{Q(x')}\right)^{2/3}$.  

The RG result \tpnine\ differs from the standard Langer
formula, since \tpnine\ involves Airy functions $\Ai$ and $\Bi$.
Note that the new variable $t$ given in \tpthree\ is a function of
$\epsilon$, and that as $\epsilon \rightarrow 0$ for fixed $x$, and $t
\rightarrow \infty$. In this limit, we can resort to the asymptotic
properties of the Airy functions $\Ai(t)$ and $\Bi(t)$ for 
$t \rightarrow \infty$, and find that $\Ai'(t) \Bi(t) \sim
-1/{2\pi}$, as $t \rightarrow \infty$. Thus, \tpnine\ recovers the
standard result 
$$C\left(t(x)\right) = C(0) (t/Q)^{1/4}. \eqn\tpten$$

However, the RG equation \tpeight\ is valid not only for relatively
large $\mu$, but also for small $\mu$. For this reason, we expect that
\tpeleven\ is a better uniformly valid approximant than the standard
Langer formula, for small and intermediate values of $t$, or for
relatively large (or not small) $\epsilon$ cases. This is verified and
can be clearly seen in Fig. \FIG\wkb{Comparison of the RG result
\tpeleven, the standard Langer formula, and the numerical solution of
equation \lthree\  for $\epsilon=0.5$ and $\epsilon=1.0$.}\wkb, where we
compare the RG result \tpeleven, the standard Langer formula, and the
exact numerical solution of equation \lthree\ for several values of
$\epsilon$. Thus, the RG results \tpeleven\ without asymptotic matching
improve upon those obtained by the standard analysis.

To conclude this section, we briefly outline the recipe to generalize
the methods for multiple-turning-point and linear
multiple-boundary-layer problems.  (For linear cases, with the help of
the transformation \ltwo\ both problems can be transformed into the
canonical form and can be treated in a unified way.)  We need only
consider the case in which $Q(x)$ in \lthree\ has multiple turning
points. Without loss of generality, we assume $Q$ has the form:
$Q(x)=f(x)\psi(x)$, where $f(x)=(x-x_1)(x-x_2)\cdots (x-x_n),\ n > 1$ is
a polynomial of $x$ with $n$ zeros $x_1 < x_2 < \cdots < x_n$, and
$\psi(x) > 0$ has no zeros.  The general strategy is first to introduce
a new independent variable $t$ defined implicitly as
$dt/\sqrt{Q/f(t)}dx/\epsilon$, where $f$ is chosen to cancel all the
zeros of $Q$.  Then we develop the straightforward perturbation series
for the resultant equation, and renormalize the integration constant to
absorb the secular divergence.  This procedure avoids performing
multiple connection formulae matching and leads to a uniformly valid
approximation.  For higher order WKB problems or linear boundary layer
problems, the generalization of the methods given here is
straightforward.

\medskip

{\bf\chapter{Switchback Problems}}
\smallskip

In previous sections, we have already seen that the RG approach not only
has conceptual, but also technical advantages compared with various
conventional methods. In this section, we will demonstrate this further,
by studying, with the aid of RG more complicated problems which involve
the so-called `switchback'.  In switchback problems, as conventionally
treated, only through subtle analysis in the course of actually solving
the problem is it possible to realize the need for, {\eg,}
unexpected order terms to make asymptotic matching consistent.

\medskip

{\bf\section{Example 1: Stokes-Oseen Caricature}}
\smallskip

A model example is a caricature of the Stokes-Oseen singular boundary
layer problem, which describes the low Reynolds number viscous flow
past a sphere of unit radius. The main result of this problem has been
presented in ref.\ {\sprl}, and in the following, we will only briefly
summarize the final results and make some additional comments.

The equation is\refmark{\hinch}
$$\frac{d^2 u}{d r^2}+\frac{2}{r}\frac{d u}{d r}+\epsilon u \frac{d
u}{d r} =0, \qquad u(1)=0, u(\infty)=1, \eqn\oseen$$
where $\epsilon$, the Reynolds number, is a small non-negative
constant.  This is a very delicate singular boundary layer problem, with 
complicated asymptotic expansions and matching,
involving unexpected orders such as $\epsilon \log(1/\epsilon)$.

Since there exists a boundary layer of thickness $\delta=O(\epsilon)$
near $r=\infty$, setting $x=\epsilon r$ transforms \oseen\ into the
following `inner' equation:  $$\frac{d^2 u}{d x^2}+\frac{2}{x}\frac{d
u}{d x}+ u \frac{d u}{d x} =0, \qquad u(x=\epsilon)=0, u(x=\infty)=1.
\eqn\stokes$$

Using RG theory, the final uniformly valid result is found to be,
to order $\lambda_1=1/e_{2}(\epsilon)$,
$$u(r;\epsilon)=1-e_{2}(\epsilon r)/e_{2}(\epsilon)
+ O\Big\{[1/e_{2}(\epsilon)]^2\Big\},\eqn\seventh$$
where the exponential integral $e_2(t)= \int_{t}^{\infty} d\rho
\rho^{-2} e^{-\rho}$, whose asymptotic expansion as $t \rightarrow 0$
is given by $e_2(t) \sim 1/t + \log{t}+(\gamma-1)-t/2+O(t^2)$ with
Euler's constant $\gamma \simeq 0.577\cdots$.

The result from asymptotic matching is given by the following
expression\rlap.\refmark{\hinch} For $r$ fixed, we have, to
$O\left(\epsilon^2 \log^2 (1/\epsilon)\right)$,
$$
u(r)\sim (1-\frac{1}{r})+\epsilon \log({1/\epsilon})(1-\frac{1}{r})
+\epsilon \big [ -\log{r} +
(1-\gamma)(1-\frac{1}{r}-\frac{\log{r}}{r})\big ], \eqn\eighth$$
while for $\rho=\epsilon r$ fixed, to $O\left(\epsilon^2
\log(1/\epsilon) \right)$,
$$u(\rho)\sim 1-\epsilon e_{2}(\rho).\eqn\ninth$$

Accordingly, examining the asymptotic result of \seventh\ in the limit
$\epsilon\rightarrow 0$, by expanding both $e_2(\epsilon r)$ and
$e_2(\epsilon)$ for $r$ fixed, and $e_2(\epsilon)$ only for
$\rho=\epsilon r$ fixed, respectively, it is found that the resulting
asymptotic solution using RG is correct to $O[\epsilon\,
\log(1/\epsilon)]$ and agrees with that obtained by asymptotic
matching.   Note that in our method, the $\epsilon \log{\epsilon}$ term
appears naturally from the asymptotic expansion of $e_2(\epsilon)$,
whereas some artistry is required to obtain this term
conventionally.    To recover the $O[\epsilon]$ term with
$(\log{r})/r$, we have to extend the RG calculation to order
$O[(1/e_2(\epsilon))^2]$.  Thus, the result to $O[\epsilon]$ given by
asymptotic matching\refmark{\hinch} is obtained from the renormalized
perturbation expansion to $O[(1/e_2(\epsilon))^2]$.

This fact may suggest that our RG result is inferior to the
conventional one.  It is important to notice, however, that neither the
asymptotic expansion (4.4) augmented with the $(\log {r})/ r$ term of
order $\epsilon$ nor (4.5) is uniformly valid in its variable $r$ or
$\rho$, respectively.  In contrast, it seems that our full result
$1-e_2(\epsilon r)/e_2(\epsilon)$ to order $\lambda_1=1/e_2(\epsilon)$
is uniformly valid as is clearly seen in Fig.\ \FIG\figio{Comparison
between the numerical solution of eq. \stokes\ for several values of
$\epsilon$, the first order RG result $1-e_2(\epsilon
r)/e_2(\epsilon)$, and two matched asymptotic expansions (one at fixed
$r$, the other at fixed $\rho\equiv r\epsilon$), as derived in ref.
\hinch.} \figio.

As discussed in the preceding paragraph, (4.3) is not an asymptotic
series in powers of $\epsilon$; thus, one might conclude that our result
is not even an asymptotic series in any sense.  Recall, however, that
the asymptotic expansion of a function is unique only when an asymptotic
sequence of functions is fixed.  The choice of the sequence is a
question of vital importance, if one wishes to have a useful asymptotic
series.  In the conventional singular perturbation methods, an
asymptotic sequence is selected by the matching conditions.  However,
there is no compelling reason to believe that the selected sequence is
practically the best asymptotic sequence (of course, it should be the
most convenient one for the matching procedure).  As we have seen, the
RG approach also produces an asymptotic sequence $\{
\lambda_{i}(\epsilon)\}$ from the requirement to satisfy the boundary
condition order by order.  Therefore, we propose the point of view that
a consistent and presumably better asymptotic expansion (starting with
$\lambda_1=1/e_2(\epsilon)$ in the present problem) may be obtained by
RG.  The standard $\epsilon$ expansion may well be an inferior
asymptotic expansion to our expansion.  In addition, the superiority to
the RG approach can also be seen from the fact that a closed expression
uniformly valid for the whole (infinite) interval has been obtained for
the problem, which is not the case for the standard asymptotic matching
method.

\medskip

{\bf\section{Example 2: Difficulty with Asymptotic Matching}}
\smallskip

To illustrate that the RG method is generally simpler to use, and
yields practically better approximants than other methods, let us next
consider a `terrible' problem whose model equation can be written
as\Ref\hinchterrible{ See ref. \hinch, pages 74 and 77.}
$$\frac{d^2 u}{d r^2}+\frac{1}{r}\frac{d u}{d r}+\alpha
\left(\frac{d u}{d r}\right)^2+
\epsilon u \frac{d u}{d r}=0, \qquad u(1)=0, u(\infty)=1, \eqn\terrible$$
where $\epsilon$ is a small non-negative constant, and $\alpha=0$ or $1$.
For $\alpha=1$, the asymptotic matching is notoriously difficult,
because an infinite number of terms must be calculated before even
the leading order can be matched successfully.  We will see how the RG
avoids such difficulties in obtaining the leading order result
uniformly valid for the entire interval $1 \leq r < \infty$.

Since there exists a boundary layer of thickness $\delta=O(\epsilon)$
near $r=\infty$, setting $x=\epsilon r$ transforms \terrible\ into the
following `inner' equation:
$$\frac{d^2 u}{d x^2}+\frac{1}{x}\frac{d u}{d x}+\alpha
\left(\frac{d u}{d x}\right)^2+
u \frac{d u}{d x}=0, \qquad u(x=\epsilon)=0, u(x=\infty)=1,
\eqn\terriblemod$$

As in other boundary layer problems, let us first look for the general
form of the solution, and then impose the required boundary conditions
to determine the constants of integration left in the solution. To do
so, we solve \stokes\ as an initial-value problem, given an initial
condition $u(x_0)=A_0$ at some arbitrary point $x=x_0$, where $A_0$ is a
finite constant.  Assuming a naive expansion
$u(x;\epsilon)=u_0(x)+\lambda_{1}(\epsilon)u_1(x)
+\lambda_{2}(\epsilon)u_2(x)+\cdots$ with initial conditions
$u_0(x_0)=A_0, u_{i}(x_0)=0, i=1,2,\cdots$, where the asymptotic
sequence $\lambda_{i}(\epsilon), i=1,2,\cdots$ are to be determined
later, we obtain
$$\frac{d^2 u_0}{d x^2}+\frac{1}{x}\frac{d u_0}{d x}+ \alpha
\left(\frac{d u_0}{d x}\right)^2+u_0 \frac{d u_0}{d x}=0. \eqn\nozero$$
The {\it finite\/} uniform solution can be guessed as
$u_0(x)=A_0$, because the uniform field should not be affected
appreciably by the distant disturbance source. Thus, the goal is to
find out the small perturbation effect on this uniform field in the
presence of a distant disturbance.

The equation for $u_1$ is
$$\frac{d^2 u_1}{d x^2}+\left(\frac{1}{x}+ A_0\right)\frac{d u_1}{d
x}=0.  \eqn\noone$$
We easily see that the equation satisfied by $u_2$ which is
significantly different from \noone\ ({\it i.e.,} with a forcing term)
appears only if $\lambda_{1}^2/\lambda_{2} =O(1)$.  We will show that
indeed the choice $\lambda_{2} = \lambda_{1}^2$ works.  The nontrivial
equation at order $\lambda_2=\lambda_1^2$ can be written as
$$\frac{d^2 u_2}{d x^2}+\left(\frac{1}{x}+ A_0\right)\frac{d u_2}{d x}=
-\alpha (\frac{d u_1}{d x})^2 - u_1 \frac{d u_1}{d x}.\eqn\notwo$$
The perturbation result is given by
$$\eqalign{
u(x)=& A_0 + \lambda(\epsilon) A_{10} \big [ e_{1}(A_0 x_0)-e_{1}(A_0 x)\big ]
+\lambda^{2}(\epsilon) \Big\{ A_2 \big [ e_{1}(A_0 x_0)-e_{1}(A_0 x)\big ]\cr
     & -\alpha \frac{1}{2} A_{10}^2 \big [ e_{1}(A_0 x_0)-e_{1}(A_0 x)\big ]^2
-A_{10}^2 A_0^{-1}\big [e_0(A_0 x)e_1(A_0 x)\cr
     & -2 e_1(2A_0 x)-e_1(A_0 x_0)e_0(A_0x)+e_0(A_0 x_0)e_1(A_0 x)\cr
     & -e_0(A_0 x_0)e_1(A_0 x_0)
      +2e_1(2A_0 x_0)\big ]\Big \}+O[\lambda^3(\epsilon)],\cr}\eqn\nothree$$
where the exponential integral $e_1(t)=\int_{t}^{\infty} \rho^{-1}
e^{-\rho}$, $\lambda_1(\epsilon)$ is already replaced by
$\lambda(\epsilon)$, and $A_{10}, A_2$ are constants of integration.
When $x_0$ is very small and $x-x_0$ is large, the divergence arises
from those terms containing $e_1(A_0 x)$ or $e_1(A_0 x_0)$, but {\it
not} $e_0(A_0 x)$ or $e_0(A_0 x_0)$. To remove the divergence from
these cross terms of $e_{0}$ and $e_{1}$, presumably both $A_0$ and
$A_1$ must be renormalized.  The renormalized perturbation result reads
$$\eqalign{
u(x)=& A(\mu) + \lambda(\epsilon) A_1(\mu) \big [ e_{1}(A \mu)-e_{1}(A x)\big ]
+\lambda^{2}(\epsilon) \Big\{ A_2 \big [ e_{1}(A \mu)-e_{1}(A x)\big ]\cr
     & -\alpha \frac{1}{2} A_{1}^2 \big [ e_{1}(A \mu)-e_{1}(A x)\big ]^2
-A_1^2 A^{-1}\big [e_0(A x)(e_1(A x)-e_1(A \mu))\cr
     & -2 e_1(2A x)
+2e_1(2A \mu)\big ]\Big \}+O[\lambda^3(\epsilon)],\cr}\eqn\nofour$$
where $A(\mu),A_1(\mu)$ are finite counterparts of
$A_0(x_0),A_{10}(x_0)$, and $\mu$ is some arbitrary length scale. The
RG equation $d u/d \mu=0$ gives
$$\frac{d A_1}{d \mu}=-\lambda(\epsilon)\alpha A_1^2 \mu^{-1}e^{-A\mu}+
O[\lambda^2(\epsilon)],\eqn\rgone$$
$$\frac{d A}{d \mu}= -\lambda(\epsilon) A_1 \mu^{-1}e^{-A\mu}
-\lambda^2(\epsilon) A_1^2 A^{-1} \mu^{-1}e^{-2A\mu}+
\lambda^2(\epsilon) A_2
\mu^{-1}e^{-A\mu}+O[\lambda^3(\epsilon)].\eqn\rgtwo$$

Now we discuss the $\alpha=0$ and $\alpha=1$ cases separately.  For
$\alpha=0$ \rgone\ suggests that $A_1$ can be treated as a constant and
there is no need to renormalize it. Solving \rgtwo\ to order $\lambda
(\epsilon)$ and setting $\mu=x$ and $x=\epsilon r$ in \nofour\ , we
obtain
$$u(r)=1-\lambda(\epsilon) A_1 e_1(\epsilon r)+\lambda^2(\epsilon),
 \eqn\extraone$$
where use is already made of the boundary condition $u(r=\infty)=1$.
Imposing $u(r=1)=0$ determines $\lambda(\epsilon) A_1=1/e_1(\epsilon)$
from which $\lambda(\epsilon)$ can be chosen as
$\lambda(\epsilon)=1/e_1(\epsilon)$, whose asymptotic expansion in the
limit $\epsilon\rightarrow 0_{+}$, is $\lambda(\epsilon)\sim
1/\log({1/\epsilon}) +\gamma/\log^2({1/\epsilon})+\cdots$, giving all
necessary orders required in the asymptotic matching. Accordingly,
$A_1=1$.  Thus, the uniformly valid asymptotic result can be written in
a single expression as
$$u(r)\sim 1- e_1(\epsilon r)/e_1(\epsilon)+O[(1/e_1(\epsilon))^2].
 \eqn\extratwo$$

For $\alpha=1$ solving \rgone\ and \rgtwo\ to order
$\lambda(\epsilon)$, we get
$$A_1(\mu)=\frac{A_1(\infty)}{1-\lambda(\epsilon)A_1(\infty)e_1(\mu)}
+O[\lambda^2(\epsilon)],\eqn\rgthree$$
$$A(\mu)=\log\Big\{1-\lambda(\epsilon)A_1(\infty)e_1(\mu)\Big\}
+A(\infty)+O[\lambda^2(\epsilon)],\eqn\rgfour$$
where $A_1(\infty), A(\infty)$ are constants of integration to be
determined by the boundary conditions.  Setting $\mu=x$ and $x=\epsilon
r$ in \nofour\, we have
$$u(r)= \log\Big\{1-\lambda(\epsilon)A_1(\infty)e_1(\epsilon r)\Big\}
+A(\infty)+O[\lambda^2(\epsilon)].\eqn\rgfive$$
Using boundary conditions $u(r=\infty)=1$ and $u(r=1)=0$ produces $A(\infty)=
1$ and $\lambda(\epsilon) A_1(\infty)=(1-1/e)/e_1(\epsilon)$.
Again we may
choose $\lambda(\epsilon)=1/e_1(\epsilon)$, and then
$A_1(\infty)=1-1/e$. Finally
the uniformly valid asymptotic result is given by
$$u(r)\sim \log\Big [ 1+(e-1)e_1(\epsilon r)/e_1(\epsilon)\Big ]
+ O[(1/e_1(\epsilon))^2].\eqn\final$$

Comparing the RG results \extratwo\ and \final\ and the corresponding
asymptotic matching results, again we find the RG results are more
accurate.


\medskip

{\bf\chapter{Reductive Perturbation Theory and RG}}
\smallskip

In previous examples, we have already mentioned the idea that
amplitude or phase equations are RG equations.  We will demonstrate
that the RG theory is a general and systematic method to derive slow
motion equations, even for those complicated problems for which no
explicit analytic zeroth-order solutions are known. In previous reports
we already discussed the one-dimensional Swift-Hohenberg
equation\refmark{\sprl} and the Burgers equation\refmark{\kawasakifest}
as renormalization group equations.  Center manifold theory can be
considered from the reductive perturbation point of view, because it
also extracts slow motion equations on the manifold.  Thus, we may
expect that the center manifold theory can also be interpreted as an
application of the renormalization approach as well.

\medskip

{\bf \section{Newell-Whitehead equation}}
\smallskip

The example we consider here is the two-dimensional Swift-Hohenberg
equation widely used as a simple model of the Rayleigh-Benard
convection\rlap,\Ref\swift{J. Swift and P. C. Hohenberg, \journal Phys.
Rev. A &15&319(77).}
$$
\frac{\partial{u}}{\partial{t}} = \epsilon u - u^3
-\left(\frac{\partial{^2}}{\partial x^2} + \frac{\partial^2}{\partial
y^2} + k^2\right)^2 u,  \eqn\sh
$$
where $\epsilon$ is a control parameter or a reduced Rayleigh number,
a measure of the degree of convective instability of the stationary
state $u=0$. For small positive $\epsilon$, the system exhibits
a supercritical bifurcation.
Since we wish to treat $\epsilon u - u^3$ as a perturbative term, to be
consistent $\epsilon u$ and $u^3$ must be of the same order.  We scale
$u$ as $\sqrt{\epsilon} u$, and denote the new $u$ with the same
symbol.  Then, the original equation reads
$$
\frac{\partial{u}}{\partial{t}} = \epsilon (u - u^3)
-\left(\frac{\partial{^2}}{\partial x^2} + \frac{\partial^2}{\partial
y^2} + k^2\right)^2 u. \eqn\shone
$$
We consider this in the whole plane for all positive $t$.  As a zeroth
order solution, we choose the roll solution along the $y$-axis:
$Ae^{ikx} +$ complex conjugate, where $A$ is a complex numerical
constant. We expand $u$ around this solution as $u = Ae^{ikx} +
\epsilon u_1 + \cdots + $ complex conjugate.  The first order
correction obeys
$$
 \frac{\partial{u_1}}{\partial{t}} + \left(\frac{\partial{^2}}{\partial
 x^2} + \frac{\partial^2}{\partial y^2} + k^2\right)^2 u_1= (1 -
3|A|^2)Ae^{ikx}.\eqn\shtwo
$$
Here, to study only the singular behavior of $u_1$, $e^{3ikx}$ and
similar non-resonant terms are ignored.  We rewrite this equation as
$$
[L_1 + L_2 + L_3 + L_4] u_1 = (1- 3|A|^2) Ae^{ikx},  \eqn\shthree
$$
where the operators are given defined as
$$
L_1 \equiv \frac{\partial}{\partial t},\;\;L_2 \equiv
\left(\frac{\partial^2}{\partial x^2} + k^2\right)^2, \;\; L_3 \equiv
2\left(\frac{\partial{^2}}{\partial{x^2} }+
k^2\right)\frac{\partial{^2}}{\partial{y^2}},\;\; L_4 \equiv
\frac{\partial{^4}}{\partial{y^4}}.\eqn\shfour
$$
We must look for space-time secular terms in the solution.  Secular
terms appear only in the special solution of the equation consistent
with the inhomogeneous term.  In order to find (space-time secular)
special solutions of \shthree\ we have only to solve $L_i u_{Si} = (1
-3|A|^2)Ae^{ikx}$ separately, and to make the linear combination of
their solutions as $\sum \mu_i u_{Si}$ with $\sum \mu_i = 1$.  This is
because all four operators $L_i$ commute, and $L_j e^{ikx} = 0$, so that
$L_iL_j u_{Si} =0$.  A trivial calculation gives
$$u_{S1} = t A (1-3|A|^2) e^{ikx}.\eqn\shfive
$$
$u_{S2}$ is governed by
$$
\left(\frac{\partial{^2}}{\partial {x^2}} + k^2\right)^2u_{S2} =
\left(\frac{\partial}{\partial x}
+ik\right)^2\left(\frac{\partial}{\partial x} -ik\right)^2 u_{S2} =
(1-3|A|^2)Ae^{ikx}.\eqn\shsix
$$
That is,
$$
\left(\frac{\partial}{\partial x} -ik\right)^2 u_{S2} =
-\frac{1}{4k^2}A(1-3|A|^2)e^{ikx}.\eqn\shseven
$$
Here we do not pay attention to inhomogeneous terms nonresonant with
the operator.  Hence, the {\it most} singular part is
$$
u_{S2} = -\frac{x^2}{8k^2}A(1-3|A|^2)e^{ikx}.\eqn\sheight
$$
Similarly, we get
$$
u_{S3} = \frac{xy^2}{8ik}A(1-3|A|^2)e^{ikx},\eqn\shnine
$$
and
$$
u_{S4} = \frac{y^4}{4!}A(1-3|A|^2)e^{ikx}.\eqn\shten
$$
In this way we get the following perturbation result,
$$
u = Ae^{ikx} + \epsilon\left(\mu_1 t - \mu_2\frac{ 
x^2}{8k^2} + \mu_3 \frac{xy^2}{8ik} +
\mu_4\frac{y^4}{4!}\right)A(1-3|A|^2) e^{ikx} +{\it c.c} + \cdots.
\eqn\sheleven$$
Here all the less singular terms ($e^{ikx}$ times $1, x, y, xy, y^2,
y^3$), higher order terms and nonsecular terms (those terms which do
not grow indefinitely far away or in the long future) are omitted.
These terms will not contribute to the final result, as shown in the
argument below.
Now, the secular terms are absorbed into the redefinition of the
amplitude $A$ as follows.  We introduce regularization points $X,Y$ and
$T$ and split, for example, $x^\alpha$ as $x^\alpha - X^\alpha+
X^\alpha$ (for some exponent $\alpha$), and absorb $X^\alpha$ into $A$.
Thus we get,
$$\eqalign{
u =& A(X,Y,T)e^{ikx} + \epsilon\Bigg(\mu_1(t-T) - \mu_2\frac{
(x^2 - X^2)}{8k^2} + \mu_3 \frac{(xy^2 -XY^2)}{8ik}\cr
& + \mu_4\frac{(y^4-Y^4)}{4!}\Bigg)A(1-3|A|^2)e^{ikx} + \cdots.\cr}
\eqn\shtwelve$$
Since $u$ should not depend on $X,Y$ or $T$, the renormalization group
equation, to $O(\epsilon)$, reads $\partial^{\alpha+\beta+\gamma} u/
\partial T^{\alpha}\partial X^{\beta} \partial Y^{\gamma} =0$ for any
positive integers $\alpha, \beta, \gamma$ with $\alpha\beta\gamma \neq
0$, where values of $\alpha, \beta, \gamma$ are chosen in such a way
that the universal
slow motion equation we are seeking is independent of any system
details. Thus, we have
$$\eqalign{
\frac{\partial A}{\partial T} - \epsilon \mu_1 A(1-3|A|^2) = 0,\cr
\frac{\partial ^2A}{\partial X^2} + \epsilon \mu_2 \frac{1}{4k^2}A
(1-3|A|^2) = 0,\cr
\frac{\partial ^3A}{\partial X\partial Y^2} - \epsilon \mu_3
\frac{1}{4ik} A (1-3|A|^2) = 0,\cr
\frac{\partial ^4A}{\partial Y^4} - \epsilon \mu_4 A (1-3|A|^2)= 0.
\cr}
\eqn\shthirteen$$
Obviously, $\mu_i$ are still almost arbitrary and must be fixed by the
auxiliary
conditions.  Therefore, to get an auxiliary condition free equation of
motion, we use $\sum \mu_i = 1$ to arrive at the following RG equation
after equating $X,Y,T$ and $x,y,t$, respectively
$$
\frac{\partial A}{\partial t} + \left(-4k^2\frac{\partial^2}{\partial
x^2} + 4ik\frac{\partial^3}{{\partial x}{\partial y^2}} +
\frac{\partial^4}{\partial y^4}\right) A = \epsilon A (1-3|A|^2).
\eqn\shfourteen$$
Thus, we have arrived at the Newell-Whitehead equation.

Let us compare this derivation with the conventional method, for which
a summary may be
found in the Appendix to the review article by Cross and
Hohenberg\rlap.\Ref\cross{M.\ C.\ Cross and P.\ C.\ Hohenberg, \journal Rev.
Mod.  Phys. &65&851(93).}  Perhaps the most notable point is that no
scaling of spatial variables like $x \rightarrow \epsilon^{1/2}x$, $y
\rightarrow \epsilon^{1/4}y$ is needed.  Furthermore, the expansion is a
straightforward one in terms of $\epsilon$ instead of
$\epsilon^{1/2}$.  That is, the result is almost automatically obtained
from the global well-definedness of the perturbation result.

If there are no spatial degrees of freedom, each step of the standard
reductive perturbation\refmark{\reductive} using the solvability
condition and that in the RG derivation above are in one-to-one
correspondence.  However, if there are spatial degrees of freedom, the
standard reductive perturbation regards the spatial derivatives as a
perturbation if the zeroth order solution is space-independent, or uses
the multiple scale analysis if the zeroth order solution is spatially
varying.  In contrast, in our RG approach, spatial and time coordinates
are treated on an equal footing, and the correct scalings of variables
are given automatically.

As the reader may have realized, kinetic equations are expected to be
derivable as slow motion equations from the BBGKY hierarchy.  For
example, the Boltzmann equation can be
derived by an RG method.  Thus we
suggest that it is a rule that slow motion equations are RG equations.

\medskip

{\bf \section{Center Manifold and RG}}

In this section, we discuss briefly the general relationship between RG
theory and center manifold theory.\refmark{\carr} In the general theory
of reduction, we wish to know the slow manifold ({\it e.g.,} inertial
manifold, center manifold) which attracts all the long-time asymptotic
solutions, and the equation of motion on the manifold.  It is well known
that the center manifold reduction and normal form theory\refmark{\carr}
have played a significant role in studying instabilities and
bifurcations encountered in dynamical systems and fluid dynamics. In
many circumstances, this approach provides a greatly simplified picture
of complicated dynamics by reducing the dimension of the system without
losing essential information concerning the instability and bifurcation.
In addition, the local dynamics on the center manifold constructed in
this way is invariant or universal, in the sense that the structure of
the reduced system is independent of specific physical models under
consideration. Thus a variety of different phenomena can have the same
type of bifurcation, belonging to the same universality class in the
parlance of RG. Although the center manifold fits in the RG picture
clearly, the general correspondence between them has not yet been
established. In certain cases such as the weakly nonlinear stability of
fluid motion, the equivalence of the method of center manifold, the
method of multiple scales, and the method of amplitude expansion has
been established explicitly by applying these methods to the derivation
of the Landau equation from the Navier-Stokes equation to the seventh
order\rlap.\Ref\fujimura{K.  Fujimura, \journal Proc.  Royal Soc. London
Series A &434&719(91).}

To illustrate the relevance of RG, let us consider the following set of
equations:
$$
\eqalign{
\frac{dx}{dt} =& f(x,y),\cr
\frac{dy}{dt} =& -y + g(x,y), \cr} \eqn\cmzero
$$
where $f$ and $g$ are higher order in the sense that $f(\lambda
x,\lambda y)$ or $g(\lambda x,\lambda y)$ is $O(\lambda^2)$ for small
$\lambda$.  Thus the variable $y$ decays quickly but $x$ does not.
Hence, the long time behavior of the system is expected to be confined
close to a local 1-manifold near the origin.  This local manifold is the
center manifold (not unique), and the long time behavior of the system
is governed by the equation of motion defined on this manifold.  Thus,
as discussed at the beginning of this section, the problem of finding a
center manifold and the equation on it is a problem of extracting slow
motion behavior of the system.  In this sense, this problem and the
general reductive perturbation can be treated in a unified fashion.
Since we are interested in the local center manifold, we may rescale the
variables as $ x \rightarrow \lambda x$ and $y \rightarrow \lambda y$,
and may assume that  $\lambda$ is small.  Therefore, instead of the
original system \cmzero, we study
$$
\eqalign{
\frac{dx}{dt} =& \lambda f(x,y), \cr
\frac{dy}{dt} =& -y + \lambda g(x,y). \cr} \eqn\cmone
$$
We assume the following formal expansions:
$$
\eqalign{
f(x,y) =& f_{20} x^2 + f_{11}xy + f_{02}y^2 + \lambda( f_{30} x^3 +
f_{21}x^2y + f_{12}xy^2 + f_{03}y^3) + \cdots\cr
g(x,y) =& g_{20} x^2 + g_{11}xy + g_{02}y^2 + \lambda( g_{30} x^3 +
g_{21}x^2y + g_{12}xy^2 + g_{03}y^3) + \cdots.\cr}\eqn\yformalll
$$
The standard approach goes as follows: Let $y=h(x)$ be the formula for
a center manifold.  Then we get the following differential equation for
$h$:
$$
-h(x) + \lambda g(x,h(x)) = \lambda h'(x) f(x,h(x)).\eqn\cmtwo
$$
This equation is usually solved by perturbation: $h(x) =  \lambda h_2 x^2
+\lambda^2 h_3 x^3 +\cdots$.  The result is
$$
y = \lambda g_{20} x^2 + \lambda^2[g_{20}(g_{11}-2f_{20}) + g_{30}]x^3 +
O[\lambda^3].\eqn\cmthree$$ The equation of motion on the center
manifold is obtained by substituting $y$ with $h(x)$ in the equation
for $dx/dt$.

Our RG program starts with the construction of a power series expansion
of the solution for \cmone\ in terms of $\lambda$ as $x = x_0 + \lambda
x_1 + \lambda^2 x_2 + \cdots$, and $y = y_0 + \lambda y_1 + \lambda^2
y_2 + \cdots$. Paquette has also pursued the same line
independently\rlap.\Ref\Paq{G.\ Paquette, private communication.}  A
lengthy but straightforward calculation gives
$$
\eqalign{
x =& A + \lambda f_{20}A^2 t  + \lambda^2 (f_{20}^2 A^3 t^2  +
f_{11}g_{20}A^3 t + f_{30}A^3 t )\cr
& +\lambda^3\{ t^3 f_{20}^3 A^4 +
\frac{5}{2}(f_{11}f_{20}g_{20}+f_{30}f_{20})A^4 t^2 \cr
& +[-2 g_{20}f_{20}f_{11} + f_{11}g_{20}g_{11} +f_{11}g_{30}
+f_{20}g_{20}^2 +f_{21}g_{20} +f_{40}]A^4 t \cr
& + CT\} + O(\lambda^4)\cr
y =& \lambda g_{20}A^2 + \lambda^2 [2g_{20}f_{20}A^3(t-1)+
g_{20}g_{11}A^3 +g_{30}A^3],
\cr} \eqn\formal
$$
where $CT$ denotes the constant terms and $A$ is the initial condition
for $x$.  Here we have discarded  all the exponentially decaying
terms.  For example, to the first order the full solution reads
$$
\eqalign{
x_1 = f_{20}A_0^2 t - f_{11}A_0B_0e^{-t} -\frac{1}{2}f_{02}B_0^2e^{-2t}
+ A_1,\cr
y_1 = g_{20}A_0^2 +g_{11}A_0B_0te^{-t} -g_{02}B_0^2e^{-2t} + B_1e^{-t},
\cr}\eqn\cmfive
$$
where $A_0,B_0,B_1$ are numerical constants dependent on the initial
data.  The exponentially decaying terms do not contribute to the
secular behavior of perturbation series.  We absorb the secular terms
proportional to the powers of $T$ into the redefined $A$ by splitting
$t$ as $ t^\alpha-T^\alpha + T^\alpha$, where $\alpha$ is an appropriate
integer.  That this can be achieved consistently must be checked order
by order.  The simplest way may be to introduce the renormalized
counterpart $A_R$ of $A$ as
$A = A_R(1+\lambda \omega_1 + \lambda^2
\omega_2 + \lambda^3 \omega_3 +\cdots)$, where $\omega_i$ are
determined to remove the powers of $T$ from the perturbation result for
$x$ after splitting $t$.  The renormalization condition can be written as
$$
\eqalign{
A_R(1+\lambda \omega_1 + \lambda^2
\omega_2 + \lambda^3 \omega_3 +\cdots) + \lambda f_{20}(1+\lambda
\omega_1 + \lambda^2
\omega_2)^2A^2t \cr
+ \lambda^2(f_{20}^2 t^2  +
f_{11}g_{20} t + f_{30}t )A^3 (1+\lambda \omega_1)^3 +
\lambda^3[\cdots] = A_R.\cr}\eqn\cmsix
$$
{}From this, order by order in powers of $\lambda$, we can fix $\omega_i$ as
$$
\eqalign{
\omega_1 =& -f_{20}At,\cr
\omega_2 =& f_{20}^2A^2t^2 -f_{11}g_{20}A^2t - f_{30}A^2t,\cr
\omega_3 =& A^3 \Bigg\{ -f_{20}^3 t^3
+\frac{5}{2}(f_{11}f_{20}g_{20}+f_{30}f_{20})t^2  \cr
& +[2g_{20}f_{20}f_{11}t - f_{11}g_{20}g_{11} -f_{11}g_{30}
-f_{20}g_{20}^2 -f_{21}g_{20} -f_{40}] t \Bigg\}.
\cr}\eqn\cmseven
$$
The renormalization group equation reads
$$
\frac{dA_R}{dT} = \frac{d}{dT}\{A_R(T)(1+\lambda \omega_1 + \lambda^2
\omega_2 + \lambda^3 \omega_3 +\cdots)\} =0.\eqn\cmeight
$$
Introducing the explicit forms of $\omega_i$ into this equation, we
experience almost miraculous cancellations of all the terms containing
powers of $t$ explicitly to have
$$
\eqalign{
\frac{dA_{R}}{dt} =& \lambda A_{R}^2f_{20} +
\lambda^2A_{R}^3(f_{11}g_{20}+f_{30})
\cr
+&\lambda^3 A_{R}^4 (f_{11}g_{30} +f_{20}g_{20}^2
+f_{21}g_{20}+f_{40} - f_{20}f_{11} g_{20} ) + O[\lambda^4],
\cr}\eqn\cmnine
$$
where $t$ is identified with $T$.  This agrees with the conventional
result.  For $y$, after renormalization, all the explicitly $t$ dependent
terms disappear to order $\lambda^2$, and
$$
y = \lambda g_{20}A_{R}^2 +\lambda^2(g_{20}g_{11} + g_{30}
-2 f_{20} g_{20})A_{R}^3+ O[\lambda^3].\eqn\cmten
$$
This also agrees with the result given above.

The formal solution \formal\ is order by order in $\lambda$ obtained
from the true solution by discarding the transcendentally small terms
in the large $t$ limit.  Notice that in $x_n$ the highest power of $t$
is $n$ (for $y_n$ it is less), so that up to a given order $n$, by
choosing $\lambda$ such that $\lambda t = 1$, we can make the
contribution of the sum of the transcendental terms (such as
$e^{-1/\lambda}$) less than any small positive number for sufficiently
large $t$.  In this way, locally up to any finite order in $\lambda$,
the series obtained as the singular (or non-decaying) terms describes
the asymptotic behavior of the system. Therefore, if the system has a
unique solution to the initial value problem (near the origin), then we
can uniquely determine these series, and they give a parametric
representation of an approximate center manifold.  In the present
context, renormalizability means that the motion on the approximate
center manifold is autonomous. The renormalization reorganizes the
expansion so that $dx/dt$ is not explicitly time-dependent.

The RG procedure given above is actually much more tedious than the
conventional approach.  However, the obtained center manifold by RG need
not be expandable in terms of $x$.  Thus, the RG method works in some
cases even when the conventional approach is not
applicable\rlap.\Ref\otcm{Y.\ Oono and E.\ Titi, unpublished (1994).}

{\bf\chapter{\bf Summary}}
\smallskip

In this paper, we have demonstrated that various singular perturbation
methods and reductive perturbation methods may be understood in a
unified fashion from the renormalization group point of view.
Amplitude equations and phase equations describing the slow motion dynamics
in nonequilibrium phenomena are RG equations. The RG method seems to be
more efficient and simpler to use than standard methods. This
is practically meaningful, because RG could yield superior approximations
without using often tedious asymptotic matching techniques.

Probably the most outstanding question is to justify mathematically the
general renormalized perturbation approach developed in this paper.
The rigorous and constructive renormalization group approaches of
Bricmont and Kupiainen and our formal perturbative approaches have almost
no common technical ground, although their philosophy is identical.
Consequently, we do not have even a hint as to how to rigorize, or
estimate the errors of our approach.

The Wilson-style RG\Ref\nato{N.\ Goldenfeld, O.\ Martin and Y.\ Oono,
{\sl Proceedings of the NATO Advanced Research Workshop on Asymptotics
Beyond All Orders}, S.\ Tanveer (ed.) (Plenum Press, 1992).} and
Bricmont and Kupiainen's related constructive renormalization group
approaches\refmark{\bricmont} can be implemented
numerically\rlap,\Ref\num{L.-Y.\ Chen and N.\ Goldenfeld, {\sl
Phys.\ Rev.\ E} (in press); M.\ Balsera, and Y.\ Oono, unpublished.}
especially with the interpolation-resampling scheme which produces a
`virtual continuum' to allow smooth scaling of any function on discrete
grids\rlap.\Ref\inter{L.\ San Martin and Y.\ Oono, submitted to
Phys.\ Rev.\ E.}

\bigskip
{\bf \ACK}

The authors are grateful to Paul Newton for valuable discussions.  YO
used material finished at the Mittag-Leffler Institute, Sweden.  The
hospitality of the Institute and useful conversations with Edriss Titi
there are gratefully acknowledged by NG and YO.  We are pleased to
acknowledge the contribution of Glenn Paquette, who participated in the
early stages of the center manifold study. LYC was in part supported
by the  National Science Foundation grant NSF-DMR-89-20538 administered
by the University of Illinois Materials Research Laboratory and in part
supported by the Institute for Theoretical Physics through National
Science Foundation Grant PHY89-04035. NG and YO gratefully acknowledge
National Science Foundation Grant NSF-DMR-93-14938 and the Mittag-Leffler
Institute for partial financial support.

\refout

\figout

\end